\documentclass{SCIS2025}
\begin{document}
%\oa
%%%%%%%%%%%%%%%%%%%%%%%%%%%%%%%%%%%%%%%%%%%%%%%%%%%%%%%
%%% Authors do not modify the information below
%%% 作者不需要修改此处信息
\ArticleType{RESEARCH PAPER}
%\SpecialTopic{}
\Year{2025}
\Month{January}
\Vol{68}
\No{1}
\DOI{}
\ArtNo{}
\ReceiveDate{}
\ReviseDate{}
\AcceptDate{}
\OnlineDate{}
\AuthorMark{}
\AuthorCitation{}
%%%%%%%%%%%%%%%%%%%%%%%%%%%%%%%%%%%%%%%%%%%%%%%%%%%%%%%

%%% title: 标题
%%%   \title{title}{title for citation}
\title{Frozen-Tag-Based Physical-Layer Authentication Against User Interference}{Frozen-Tag-Based Physical Layer Authentication Against User Interference}

%%% Corresponding author: 通信作者
%%%   \author[number]{Full name}{{email@xxx.com}}
%%% General author: 一般作者
%%%   \author[number]{Full name}{}
%%% Equal Contribution: 同等贡献作者
%%%   \author[number\dag]{Full name}{}
\author[1]{Lei YAO}{}
\author[1]{Boxiang HE}{boxianghe1@bjtu.edu.cn}
%\author[1]{Weiyu CHEN}{}
\author[1]{Shilian WANG}{wangsl@nudt.edu.cn} 
\author[2]{Ning XIE}{}
%%% Authors' contribution. 同等贡献声明
%\contributions{These authors contributed equally to this work.}

%%% Address. 地址
%%%   \address[number]{Affiliation, City Postcode, Country}
\address[1]{College of Electronic Science and Technology, National University of Defense Technology, Changsha 410073, China}
\address[2]{College of Electronics and Information Engineering, Shenzhen University, Shenzhen 518060, China}

%%% Abstract. 摘要
\abstract{Tag-based physical layer authentication (PLA) has garnered significant attention due to its low complexity and enhanced security. However, existing PLA schemes encounter two challenges. First, unintended user interference, which overlaps with the authentication signal, corrupts the tag and degrades authentication performance. Second, the vulnerability introduced by direct embedding of the raw tag exposes the tag to the adversary and degrades the security. To address these challenges, this paper proposes a novel frozen-tag-based PLA framework. Different from typical schemes that directly embed the uncoded tag into the signal, a well-designed frozen tag is inserted for authentication, where the frozen tag is generated based on the concept of polar codes with the anchor information as information bits and raw tags as frozen bits. Accordingly, the proposed PLA framework offers two principal advantages. First, the authentication performance is improved since the legitimate receiver can decode the frozen tag and mitigate unintended user interference. Second, the authentication process becomes indecipherable to the illegitimate receiver due to the concealment of the raw tags. Furthermore, we conduct a comprehensive analysis of the proposed framework in terms of robustness, security, and compatibility. Specifically, the security analysis demonstrates that an eavesdropper faces a high error probability when locating frozen tags, and accumulates noise power that increases with the length of the frozen tag  during the estimation of the raw tag. Regarding robustness and compatibility, we derive a union bound on the detection probability and an upper bound on the bit error rate of the message, respectively. Theoretical analysis and simulation demonstrate that the proposed frozen-tag-based PLA framework not only enhances the detection performance but also significantly degrades Eve's capability to estimate the raw tags.}

%%% Keywords. 关键词

\maketitle

%%%%%%%%%%%%%%%%%%%%%%%%%%%%%%%%%%%%%%%%%%%%%%%%%%%%%%%
%%% The main text. 正文部分
%%%%%%%%%%%%%%%%%%%%%%%%%%%%%%%%%%%%%%%%%%%%%%%%%%%%%%%
\vspace*{-0.7cm}
\section{Introduction}
\vspace*{-0.2cm}
The advancement of fifth-generation (5G) and future sixth-generation (6G) communications are driving their expansion into critical areas such as massive Internet of Things (IoT)\cite{iot,6G}, industrial automation\cite{indusIOT}, and Internet of Vehicles (IoV)\cite{IOV}. This progress has led to a significant increase in the number of user devices and the volume of data\cite{IOTT}. Thus, traditional upper-layer authentication (ULA) protocols are increasingly challenged due to high computational complexity, high communication overhead, and high latency\cite{ULA}. The limitations are more significant in  intensive and low-latency scenarios, such as real-time signaling for autonomous vehicles or instant control in industrial sensor networks, where the significant security threat and management overhead imposed by ULA not only creates severe performance bottlenecks but may introduce new vulnerabilities\cite{XieSurvey,XieTSP}. Thus, it is crucial to explore lightweight and secure authentication mechanisms at the physical layer, i.e., physical layer authentication (PLA).

PLA aims to verify the identity of communication entities by unique and measurable physical properties in communication signals, such as channel characteristics\cite{Channel1,Channel2,Channel3}, radio frequency fingerprints\cite{device1}, or embedded tags\cite{PLA2008,XieSlope}. Compared with ULA, it offers the following advantages. First, PLA exhibits low overhead and latency. Specifically, PLA typically eliminates the need for complex encryption by leveraging inherent physical characteristics, such as channel response or tag, for identity verification. Thus, it significantly reduces computational complexity and authentication latency. Second, PLA can achieve information-theoretic security\cite{ITS}. Unlike ULA, the security of PLA is rooted in the randomness, uniqueness, and reciprocity of the wireless channel. Thus, it is difficult for the adversary to impersonate a legitimate identity or deduce secret keys.

PLA can be categorized into passive and active schemes, where the former exploits physical characteristics in communication, such as channel response \cite{emc,CIR1} and frequency offset\cite{CFO1,CFO2}, for authentication. Passive schemes are fundamentally limited by the stability of wireless channel characteristics and the associated need for highly accurate measurement. Furthermore, such schemes are inherently vulnerable to eavesdropping attacks, which significantly undermines the security of the system. In contrast, active schemes, i.e., tag-based PLA schemes, can offer greater flexibility and enhanced resilience against adversaries by well-designed authentication tags.

Tag-based PLA is performed by embedding a tag into the message. The first work on tag-based PLA is achieved by directly superimposing authentication tags onto the message\cite{PLA2008}. Since the power of the tag is much lower than that of the message, the authentication signal remains well concealed from eavesdroppers while maintaining good compatibility with unaware receivers. Building upon this, subsequent efforts focused on achieving high security, high robustness and low complexity. For instance, a slope authentication scheme is proposed \cite{XieSlope}, which eliminates the need for complex preprocessing such as channel estimation. A blind PLA scheme is proposed\cite{BTP}, where the receiver can perform authentication without the knowledge of the authentication parameters, thereby reducing the complexity. By combining the tags with the challenge-response (CR) authentication mechanism\cite{CR}, a CR-based hybrid scheme is introduced\cite{CRH}. This scheme adjusts the transmission power of the authentication signal according to channel fading, thereby significantly improving the robustness of authentication. To resist the spoofing and replaying attacks, a Gaussian tag-based PLA scheme is proposed by using weighted fractional Fourier transform\cite{WFRFT}. Further innovations have introduced schemes operating in diverse domains, such as asynchronous tag-based PLA\cite{ATBS}, which enhances security and compatibility through an artificial delay, and phase tag-based PLA\cite{PhaPLA}, where the tag is superimposed onto the phase of the signal.

The aforementioned existing schemes are largely confined to single-user or idealized multiuser scenarios. In practical multiuser communication, such as IoT and IoV, signals from multiple users are superimposed at the receiver. Due to the imperfections in multiuser detection techniques, such as imperfect successive interference cancellation in non-orthogonal multiple access systems\cite{Interf2}, residual interference from the detection of quasi-orthogonal pilot\cite{emc}, and other multiuser detection techniques\cite{Interf1}, the authentication signal of the target user is disturbed and even obscured by residual signals from other users. For simplicity, this effect is referred to as user interference in this paper.

Unfortunately, existing PLA schemes, which can be categorized as uncoded tag-based schemes due to the fact that they directly insert the tag into the message, suffer from two significant challenges. First, the authentication performance degrades significantly under user interference. Specifically, uncoded tag-based schemes are vulnerable to user interference since it can severely distort the low-power authentication tags and lead to a degradation in authentication performance. Second, uncoded tag-based schemes are vulnerable to eavesdropping attacks. Specifically, existing schemes directly superimpose raw tags, exposing them to eavesdroppers and thereby posing a serious security threat.\cite{MulOb}.

To address the aforementioned challenges, this paper proposes a novel frozen-tag-based PLA framework by carefully designed authentication tags. Specifically, the main contributions of this paper are:
\begin{itemize}
	\item We propose a novel PLA framework based on frozen tag, where the frozen tag is the coded version of the raw tag. Unlike traditional schemes that directly superimpose raw tags onto message, the proposed framework constructs coded frozen tags by the concept of the polar code. This design significantly enhances robustness against user interference and raises the difficulty for adversaries to extract authentication tags. Thus, the security and robustness in the proposed scheme are significantly improved.
	\item The closed-form expressions of the proposed framework are derived in terms of robustness, security, and compatibility. In particular, regarding security, we derive the probability that Eve correctly classifies the position of the tag, as well as the power of the accumulated noise when Eve estimates the raw tag. Our theoretical analysis reveals that it is difficult for Eve to launch eavesdropping attacks due to the extremely low correct classification probability and the strong accumulated noise during the estimation of the raw tag. Moreover, for robustness, a union bound on the detection probability is derived. For compatibility, an upper bound of bit error rate (BER) on the message decoding is provided by modeling the tag insertion and the wireless channel as a cascaded channel.
	\item The numerical results demonstrate that the proposed PLA framework outperforms traditional uncoded tag-based schemes in terms of robustness to user interference and receiver noise. Moreover, the proposed framework can further increase the detection probability by extending the length of the frozen tags. Regarding security, the proposed framework makes it difficult for an eavesdropper to estimate the tag. Specifically, Eve faces challenges in successfully estimating the position of the frozen tags and the raw tags. In terms of compatibility, the proposed framework exhibits high flexibility since it can control the BER of the tagged signal by adjusting the length of the frozen tags.
\end{itemize}

The remainder of this paper is organized as follows. Section \ref{2} introduces the system and typical framework, and the limitations of existing framework are stated. The frozen-tag-based PLA framework and the carefully designed modules are proposed in Section \ref{3}, and the performance of the proposed framework in terms of robustness, security, and compatibility is analyzed in Section \ref{4}. In Section \ref{5}, the numerical results are carefully presented. Section \ref{6} concludes this paper.
\begin{figure}[!t]
	\setlength{\abovecaptionskip}{0pt}
	\centering
	\captionsetup{font={scriptsize}}
	\includegraphics [width=4in]{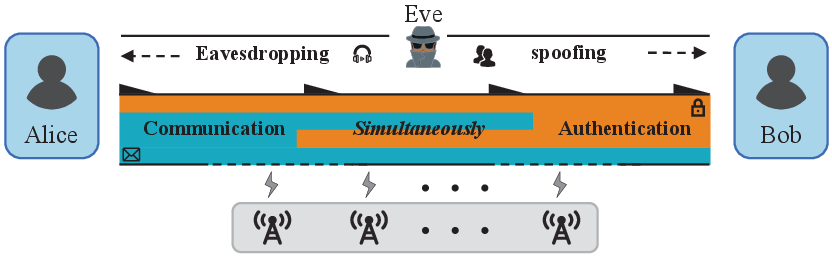}
	\caption{A typical authentication scenario, where Alice sends a signal to Bob for both authentication and communication tasks under Eve's eavesdropping in a multiuser communication.}
	\label{SysM}
	\vspace*{-0.3cm}
\end{figure}

\emph{Notation}: Throughout this paper, scalars and vectors are denoted by lower-case italic letters $x$ and bold lower-case italic letters $\boldsymbol{x}$, respectively. The operator $\Re\{\cdot\}$ denotes the real part. $\boldsymbol{x}_{i}$ denotes the $i$-th bit of signal $\boldsymbol{x}$.
$\boldsymbol{x}_{\mathcal{A}}$ denotes the set of elements in vector $\boldsymbol{x}$  indexed by $\mathcal{A}$. $[\cdot]^*$, $[\cdot]^{\mathsf{T}}$, and $[\cdot]^{\dag}$ denote the conjugate, the transpose, and the Hermitian, respectively. $[K]$ denotes the set of integers $\left\{1,2,\cdots,K\right\}$. The operator $\mathbb{E}(\cdot)$ denotes the expectation. $\boldsymbol{x}\sim \mathcal{CN}\left(\boldsymbol{u},\bf{\Sigma}\right)$ represents the circularly symmetric complex Gaussian (CSCG) random vector $\boldsymbol{x}$ with the mean $\boldsymbol{u}$ and the covariance matrix $\bf{\Sigma}$. $\text{exp}(\cdot)$ is the exponential function. $\mathcal{A}^\text{c}$ is the complementary set of the set $\mathcal{A}$. $\mathbb{C}^{x\times y}$ is the space of $x \times y$ complex-valued matrices. $\mathbb{F}_2^{x\times y}$ denotes the finite field space of 0 and 1 with dimension $x\times y$. $\mathbf{I}_m$ is the identity matrix of order $m$.

\vspace*{-0.5cm}
\section{System Model and Typical Framework}\label{2}
\vspace*{-0.2cm}

In this section, we first describe the system model of the tag-based PLA. Then, the typical tag-based authentication framework and its limitations are briefly reviewed.

\vspace*{-0.3cm}
\subsection{System Model}
\vspace*{-0.2cm}
We consider a typical authentication scenario with user interference, as depicted in Figure \ref{SysM}, where Alice sends a signal to Bob for both authentication and communication tasks, while other $K$ users simultaneously transmit signals to Bob for communication task. Since Bob receives a superposition $\boldsymbol{y}$ of multiple signals, it is challenging to extract precise authentication signal through multiuser detection techniques. Thus, the imperfect multiuser detection leads to the user interference, i.e.
\vspace*{-0.2cm}
\begin{align}\label{111}
	{\boldsymbol{y}} = {h}{{\boldsymbol{x}}} +\!\!\!\!\!\!\!\underbrace{\sum\limits_{k = 1}^{K} {\alpha _k}{{\boldsymbol{x}}_k}}_{\text{User interference}~\boldsymbol{I}} \!\!\!\!\!\!\!+ {{\boldsymbol{w}}_{\text{B}}},
\end{align}
where $h$ is a block-fading channel between Alice and Bob; $\boldsymbol{x}$ is the authentication signal generated by the secret key $\boldsymbol{k}$; $\boldsymbol{x}_k,k\in[K]$, is the interference from the $k$-th user; $\alpha_k$ is the interference weight; $\boldsymbol{w}_{\text{B}}\sim \mathcal{CN}\left(\boldsymbol{0}, \sigma^2\mathbf{I}_L\right)$ is the receiver noise at Bob\cite{emc,Interf1,Interf2}. Bob makes the authentication decision between the following hypotheses:
\vspace*{-0.2cm}
\begin{align}
	\begin{array}{l}
		{\text{H}_0}:{\text{The\enspace received\enspace signal\enspace is\enspace  illegitimate}},\\
		{\text{H}_1}:{\text{The\enspace received\enspace signal\enspace is\enspace legitimate}}.
	\end{array}
\end{align}
Note that the probability of accepting $\text{H}_1$ when $\text{H}_1$ is true is the detection probability, denoted by $P_\text{D}$, while the probability of incorrectly rejecting $\text{H}_0$ when $\text{H}_0$ is true is the false alarm probability, denoted by $P_\text{FA}$. Furthermore, a potential adversary, Eve, can launch eavesdropping and spoofing attacks. Although Eve is unaware of the secret key, he possesses powerful computational ability. 

\begin{figure}[!t]
	\setlength{\abovecaptionskip}{0pt}
	\centering
	\captionsetup{font={scriptsize}}
	\includegraphics [width=5in]{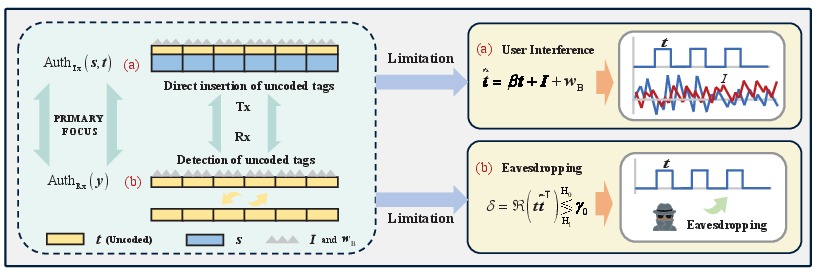}
	\caption{Typical uncoded tag-based PLA schemes, such as the superimposed tag method, suffer from two limitations: (a) performance degradation under user interference, and (b) vulnerability to eavesdropping attacks.}
	\label{limi}
	\vspace*{-0.3cm}
\end{figure}

\vspace*{-0.2cm}
\subsection{Overview of Typical Uncoded Tag-Based PLA Framework}
\vspace*{-0.2cm}

To clearly demonstrate the superiority of the proposed framework against user interference, the typical uncoded tag-based PLA framework is briefly reviewed in this subsection. In the typical framework, Alice covertly embeds the uncoded tag, i.e., the raw tag, into the message, while Bob attempts to detect the tag from the received signal for authentication.

Specifically, Alice embeds the tag $\boldsymbol{t}$ into the message $\boldsymbol{s}$ to obtain the authentication signal, i.e.
%\vspace*{-0.2cm}
\begin{align}
	{\boldsymbol{x}} = {\text{Aut}}{{\text{h}}_{{\text{Tx}}}}\left( {{{\boldsymbol{s}}},{\boldsymbol{t}}} \right),
\end{align}
where ${\text{Aut}}{{\text{h}}_{{\text{Tx}}}}\left(\cdot\right)$ denotes the authentication signal generation function that governs how the tag is inserted. There are two typical forms of ${\text{Aut}}{{\text{h}}_{{\text{Tx}}}}\left(\cdot\right)$, including superimposing the tag onto the message with low power, i.e., the superimposed tag method\cite{PLA2008}, and replacing the message with the tag, i.e., the replaced tag method\cite{XiePSA}.

At the receiver, Bob obtains the estimation of the tag $\hat{\boldsymbol{t}}$ from the received signal $\boldsymbol{y}$, i.e.
\begin{align}
	\hat {\boldsymbol{t}} = {\text{Aut}}{{\text{h}}_{{\text{Rx}}}}\left( {\boldsymbol{y}} \right),
\end{align}
where ${\text{Aut}}{{\text{h}}_{{\text{Rx}}}}\left(  \cdot  \right)$ is the tag estimation function paired with ${\text{Aut}}{{\text{h}}_{{\text{Tx}}}}\left(\cdot\right)$.  The authentication is performed by detecting the uncoded tag from the estimated tag, i.e.
\vspace*{-0.2cm}
\begin{align}
	\delta  = \Re\left(\boldsymbol{t}\hat{\boldsymbol{t}}^\mathsf{T}\right) \underset{\text{H}_1}{\overset{\text{H}_0}{\lessgtr}}  \gamma_0,
\end{align}
where $\gamma_0$ is the detection threshold. It is worth noting that most typical PLA frameworks focus on designing the paired ${\text{Aut}}{{\text{h}}_{{\text{Tx}}}}\left(  \cdot  \right)$ and ${\text{Aut}}{{\text{h}}_{{\text{Rx}}}}\left(  \cdot  \right)$ functions. Moreover, they perform authentication by detecting uncoded tags. However, the typical schemes have two limitations, as depicted in Figure\ref{limi}:
\begin{itemize}
	\item  First, the robustness of uncoded tag-based schemes is degraded under user interference. Specifically, typical approaches estimate the tag by subtracting the estimated message $\boldsymbol{s}$
	from the received signal $\boldsymbol{y}$. However, even when the message is perfectly estimated, residual user interference persists in the estimated tag. We take the superimposed tag method as an example, where the authentication signal is given by $\boldsymbol{x}=\rho_\text{s}\boldsymbol{s}+\rho_\text{t}\boldsymbol{t}$ with $\rho_\text{s}$ and $\rho_\text{t}$ being the power allocations. From (\ref{111}), the tag can be estimated by  removing message $\boldsymbol{s}$ from the received signal $\boldsymbol{y}$, i.e.
	\vspace*{-0.2cm}
	\begin{align}\label{666}
		\hat{\boldsymbol{t}} = \rho_\text{t}h\boldsymbol{t}+\boldsymbol{I} + {{\boldsymbol{w}}_{\text{B}}}.
	\end{align}
	As can be observed from (\ref{666}), unintended user interference perturbs the raw tag $\boldsymbol{t}$ and leads to a degradation in the detection performance.
	\item Second, the typical schemes are susceptible to eavesdropping attacks. Specifically, typical schemes directly embed the raw tags into the messages. Thus, the adversary can easily estimate the tags due to the direct accessibility of the uncoded tag.
\end{itemize}

In contrast to the conventional focus on optimizing the paired ${\text{Aut}}{{\text{h}}_{{\text{Tx}}}}\left(  \cdot  \right)$ and ${\text{Aut}}{{\text{h}}_{{\text{Rx}}}}\left(  \cdot  \right)$ functions, this paper identifies the design of the tag $\boldsymbol{t}$ as a pivotal yet neglected dimension for enhancing authentication performance under user interference and eavesdropping attacks. Specifically, we propose a novel PLA paradigm that shifts from the  typical uncoded tag to the novel coded tag. First, user interference is overcome by designing a frozen tag, i.e., the coded tag, via polar encoding of both the raw tag and the authentication information. Second, in the generation of the frozen tags, the raw tags are concealed within the frozen bits. In this way, the raw tags facilitate the decoding of the authentication information and inherently resist eavesdropping attacks.

\begin{figure*}[t]
	\centering
	\includegraphics[width=1\linewidth]{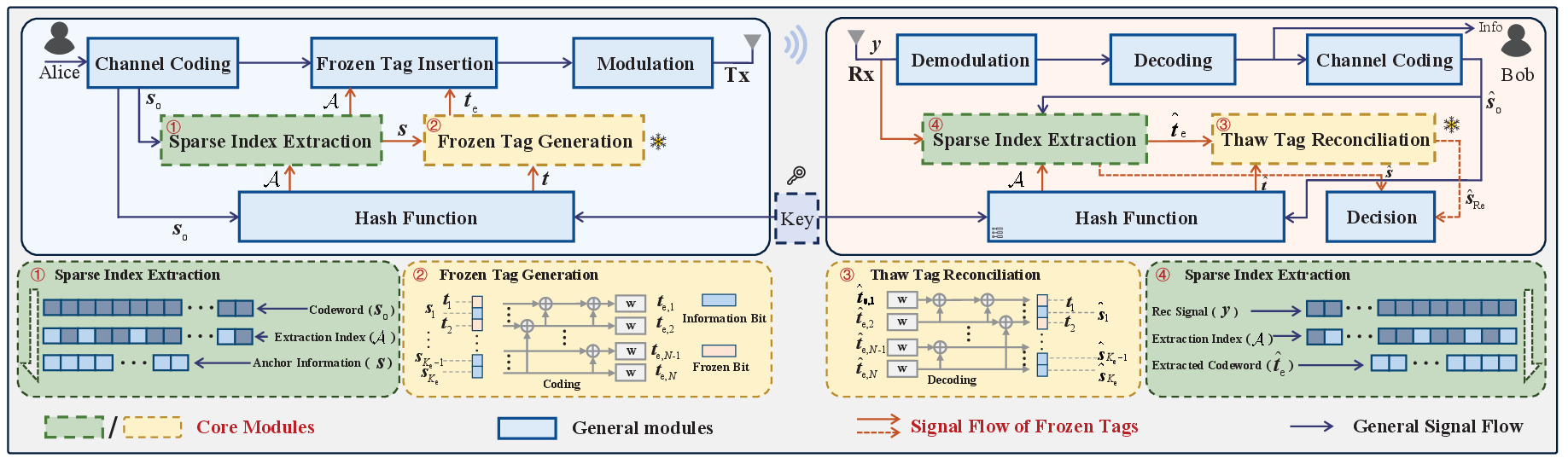} 
	\captionsetup{font=scriptsize}
	\caption{Proposed PLA framework for multiuser communication, where the symmetric Frozen Tag Generation and Thaw Tag Reconciliation modules (Yellow), and paired Sparse Index Extraction modules (Green) at the transceiver are specially designed to counter the user interference and the eavesdropping attacks.}
	\label{MultiUserFramework}
	\vspace*{-0.3cm}
\end{figure*}

\vspace*{-0.5cm}
\section{Proposed Frozen-Tag-Based PLA Framework}\label{3}
\vspace*{-0.2cm}
Compared with conventional framework, we carefully design the paired Frozen Tag Generation (FTG) and Thaw Tag Reconciliation (TTR) modules. Here, a piece of  anchor information is specified at the transmitter and is estimated by the receiver to construct the test statistic, while the raw tag is concealed during transmission and facilitates the decoding of the anchor information as side information. 
%Specifically, the FTG module generates the coded frozen tags by placing anchor information on the information bits of the polar code and the raw tags on the frozen bits, respectively. Correspondingly, the TTR module utilizes the raw tags as side information to estimate the anchor information. Moreover, thanks to the paired Sparse Index Extraction (SIE) Modules, the proposed framework demonstrates high compatibility. 
To clarify the proposed framework, the key terms are defined in Table \ref{terminology}.

\begin{table}[t]
	\centering
	\caption{Key Terminologies in the Proposed Frozen-Tag-Based PLA Framework}
	\label{terminology}
	\footnotesize  % ???????
	\begin{tabularx}{\linewidth}{@{} l @{\quad} X @{}}  % ?? >{\bfseries}
		\toprule
		\textbf{Term} & \textbf{Description} \\  % ?????
		\midrule
		Raw Tag $\boldsymbol{t}$ & The original authentication information generated by (\ref{999}). \\ 
		Frozen Tag $\boldsymbol{t}_\text{e}$ & The encoded version of the raw tag, constructed by (\ref{III}), (\ref{FFF}) and (\ref{Gn}). \\ 
		Message $\boldsymbol{s}_\text{o}$ & The communication signal. \\ 
		Anchor Information $\boldsymbol{s}$ & Public message selected from $\boldsymbol{s}_\text{o}$ to construct the test statistic. \\ 
		\bottomrule
	\end{tabularx}
\vspace*{-0.2cm}
\end{table}

The proposed authentication framework comprises two phases: the preparation and the authentication phases. In the preparation phase, Bob and Alice estimate the channel  $\hat{h}$. In the authentication phase, Alice sends a tagged signal to Bob for authentication. In the rest of this section, we first introduce the overview of the proposed framework. Then,  the key modules are detailed.
\vspace*{-0.2cm}
\subsection{Overview of Proposed Frozen-Tag-Based PLA Framework}
\vspace*{-0.2cm}
The overview of the proposed framework is illustrated in Figure \ref{MultiUserFramework}. At the transmitter, a frozen tag, which is constructed by both the raw tag and the anchor information, is embedded into the message, while at the receiver, the anchor information is estimated to construct the test statistic.

Specifically, Alice sends a signal $\boldsymbol{x}$ to Bob for authentication, where $\boldsymbol{x}=\boldsymbol{x}_\text{Mod}/\hat{h}$ and $\boldsymbol{x}_\text{Mod}$ is the modulated version of the tagged signal ${{\boldsymbol{s}}_{\text{t}}}\in \mathbb{F}_2^{{1\times N}}$. The tagged signal is obtained by 
replacing a selected subset $\cal A$ of the message ${{\boldsymbol{s}}_{\text{o}}} \in \mathbb{F}_2^{1 \times N}$ with the frozen tag ${{\boldsymbol{t}}_{\text{e}}} \in \mathbb{F}_2^{{1\times N_{\text{e}}}}$, i.e.
\begin{align}\label{777}
	\boldsymbol{s}_{\text{t},\mathcal{A}}=\boldsymbol{t}_\text{e},
	\vspace*{-0.5cm}
\end{align}
and
\begin{align}
	\boldsymbol{s}_{\text{t},\mathcal{A}^{\text{c}}}=\boldsymbol{s}_{\text{o},\mathcal{A}^\text{c}},
\end{align}
	where ${\cal A}^{\text{c}}=[N]\setminus{\cal A}$ is the complementary index set, and $[N]$ denotes the full index set of the message. The index set of the replacement positions $\cal{A}$, which satisfies $|{\cal A}| = {N_\text{e}}$, is determined by a secret key $\boldsymbol{k}$ through a one-way hash function $\text{Gen}_{\text{Pos}}\left(\cdot\right)$, i.e.
	\begin{align}\label{888}
		{\cal A} = {\text{Ge}}{{\text{n}}_{{\text{Pos}}}}\left( {{{\boldsymbol{s}}_{\text{o}}},{\boldsymbol{k}}} \right),
	\end{align}
	To ensure that the message $\boldsymbol{s}_{\text{o}}$ can be recovered at the receiver, the length of the frozen tag is much smaller than that of the message, i.e., $N_\text{e}\ll N$. Moreover, the frozen tag $\boldsymbol{t}_\text{e}$ is generated by the FTG module, which takes two inputs: the anchor information $\boldsymbol{s}$ and the raw tag $\boldsymbol{t}$. The anchor information $\boldsymbol{s}$ is specified by the first $K_{\text{e}}$ bits of the message indexed by $\mathcal{A}$, i.e., $\boldsymbol{s}=\boldsymbol{s}_{\text{o},\mathcal{A}^\prime}$ with $\mathcal{A}^\prime=\mathcal{A}_{[K_\text{e}]}$, and is then estimated at the receiver to construct the test statistic. The raw tag $\boldsymbol{t}$ is derived from the message $\boldsymbol{s}_{\text{o}}$ and the secret key $\boldsymbol{k}$ using another one-way hash function $\text{Gen}_{\text{Tag}}(\cdot)$, i.e.
	\vspace*{-0.2cm}
	\begin{align}\label{999}
		\boldsymbol{t} = \text{Gen}_{\text{Tag}}\left( \boldsymbol{s}_{\text{o}}, \boldsymbol{k} \right).
	\end{align}

	From (\ref{111}), the received signal at Bob is the superposition of the authentication signal and residual user interference, i.e.
	\vspace*{-0.2cm}
	\begin{align}
		{\boldsymbol{y}} = {{\boldsymbol{x}}_{{\text{Mod}}}} + \boldsymbol{I} + {{\boldsymbol{w}}_{\text{B}}}. 
	\end{align}
	Based on $\boldsymbol{y}$, Bob constructs the test statistic by comparing two estimations of the anchor information: one derived from the decoded message and the other obtained by the TTR module. Specifically, the test statistic is constructed as
	\vspace*{-0.2cm}
	\begin{align}\label{TS}
		\delta  = \sum\limits_{i = 1}^{{K_{\text{e}}}} {\left( {1 - |{{{\boldsymbol{\hat s}}}_{{\text{Re,}}i}} - {{{\boldsymbol{\hat s}}}_i}|} \right)}\underset{\text{H}_1}{\overset{\text{H}_0}{\lessgtr}}  \gamma_0,
	\end{align}
	where $\gamma_0$ is the detection threshold. 
	%$\hat{\boldsymbol{s}}$ is the anchor information estimated directly from the decoded message, and $\hat{\boldsymbol{s}}_{\text{Re}}$ denotes the reconciled message estimated by the TTR module. 
	The first estimation ${\boldsymbol{\hat s}}$ is obtained directly from the decoded message ${\boldsymbol{\hat s}_\text{o}}$, which is recovered through demodulation, decoding, and re-encoding of $\boldsymbol{y}$. With the estimated message $\hat{\boldsymbol{s}}_\text{o}$ and the secret key $\boldsymbol{k}$, the set $\hat{\cal{A}}$ can be estimated using (\ref{888}). The anchor information is then extracted as ${\boldsymbol{\hat s}} = {{\boldsymbol{\hat s}}_{{\text{o}},\hat{\mathcal{A}}_{[K_\text{e}]}}}$.
	The second estimation, i.e., the reconciled message $\hat{\boldsymbol{s}}_{\text{Re}}$, can be obtained by the TTR module with both the estimated raw tag $\hat{\boldsymbol{t}}$ and  the noisy observation of frozen tag ${\boldsymbol{\hat t}}_{\text{e}}$ being the inputs.  The raw tag $\hat{\boldsymbol{t}}$ can be estimated using (\ref{999}) with the message $\hat{\boldsymbol{s}}_\text{o}$. Moreover, since the frozen tag is directly inserted into the message based on (\ref{777}), ${\boldsymbol{\hat t}}_{\text{e}}$ can be retrieved from the received signal, i.e.
	%\vspace*{-0.2cm}
	\begin{align}
		{{\boldsymbol{\hat t}}_{\text{e}}} = {\boldsymbol{y}}_{\hat{\mathcal{A}}}.
	\end{align}
	\begin{remark}
		In contrast to typical uncoded tag-based PLA schemes that directly transmit the raw tags, the proposed framework transmits a processed version of the raw tag, i.e., the frozen tag in (\ref{777}). This design prevents Eve from directly eavesdropping on the raw tag. Furthermore, without the knowledge of the secret key, it is also difficult for Eve to estimate the raw tag from the noisy observation of the frozen tag, which will be elaborated in Section \ref{SecAna}.
	\end{remark}
	
	In the following, we provide a detailed description of the key modules.
	
	%\begin{remark}
	%	In contrast to typical PLA schemes, which directly superimpose or insert the raw tags into the message\cite{PLA2008,CRH,XiePSA}, the proposed framework generates frozen tags using both the raw tags and the anchor information. In terms of robustness, the raw tags facilitate the decoding of the anchor information, thereby improving robustness. From a security perspective, the proposed framework offers two advantages. On one hand, since the anchor information is part of the message, the frozen tags are message-dependent.  This prevents an eavesdropper from launching a spoofing attack using the intercepted frozen tags, as these tags are not applicable to an arbitrary message. On the other hand, it is difficult for an eavesdropper to intercept the raw tags, given that the frozen bits, where the raw tags are placed, are of low reliability, making them difficult to decode correctly.
	%\end{remark}
	
	\begin{figure*}[t!]
		\centering
		\begin{minipage}[t]{0.49\linewidth}
			\centering
			\includegraphics[width=3in]{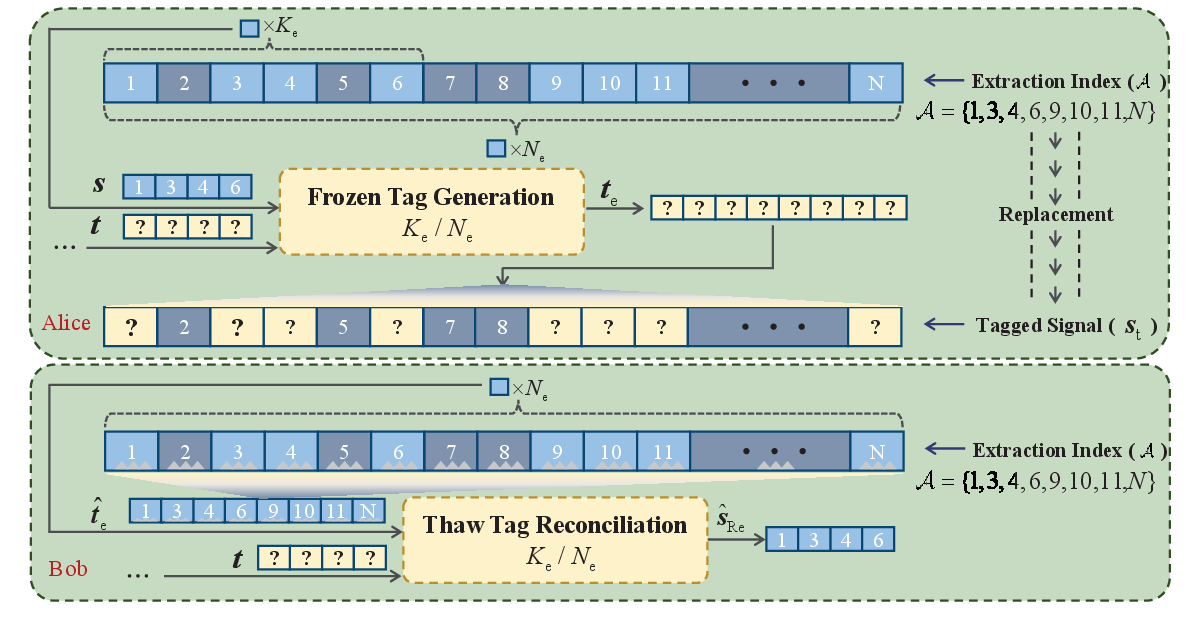}
			\caption{Detailed signal flow of the paired Sparse Index Extraction Modules with an example, where the upper branch (Tx) generates the information $\boldsymbol{s}$ and the index set $\mathcal{A}$ for the insertion of frozen tags, while the lower branch (Rx) retrieves the noisy frozen tag from the received signal.}
			\label{SIE}
		\end{minipage}
		\hfill
		\begin{minipage}[t]{0.49\linewidth}
			\centering
			\includegraphics[width=3in]{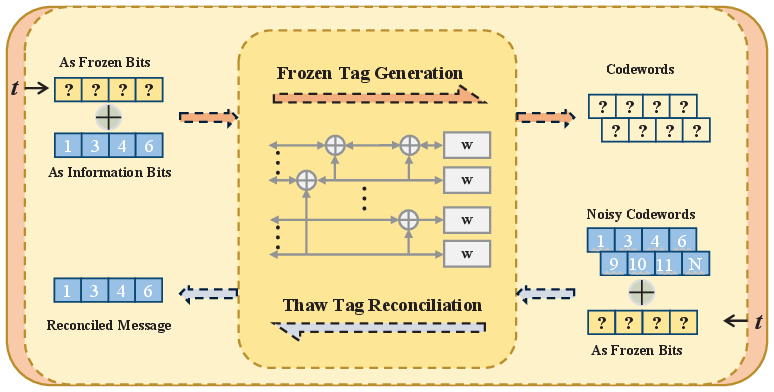}
			\caption{Detailed signal flow of the paired Frozen Tag Generation (Red Arrows) and Thaw Tag Reconciliation (Blue Arrows) modules, where the receiver employs the same tag $\boldsymbol{t}$ as the transmitter to facilitate decoding and ensure correct reconciliation.}
			\label{PolarModule}
		\end{minipage}
		\vspace*{-0.3cm}
	\end{figure*}
	
\vspace*{-0.2cm}	
	\subsection{Paired Sparse Index Extraction Modules}\label{PSIEMs}
\vspace*{-0.2cm}
	The paired SIE modules aim to enhance the compatibility of the proposed framework by randomly replacing a subset of the message with frozen tags. In the proposed framework, both the transmitter and receiver are equipped with a SIE module. Specifically, at the transmitter, the module extracts the anchor information and determines the positions of frozen tags based on the raw tags, while at the receiver, it extracts the noisy observation of the frozen tags from the received signal and generates the estimated anchor information. The detailed signal flow within the SIE module is illustrated in Figure \ref{SIE}. Here, an example is provided below to clarify the purpose of the SIE module.
	
	At the transmitter, Alice first extracts the anchor information $\boldsymbol{s}$. Specifically, the SIE module obtains the index set $\mathcal{A}=\left\{1,3,4,6,9,10,11,N\right\}$ with $|\mathcal{A}|=8$ from the Hash Function module. Assuming that the coding rate for the FTG module is $K_\text{e}/N_\text{e}=1/2$, the length of the anchor information $\boldsymbol{s}$ is given by $K_\text{e}=4$. Then, Alice extracts the anchor information $\boldsymbol{s}$ from the raw message $\boldsymbol{s}_\text{o}$ with the first $K_\text{e}$ indices in $\mathcal{A}$, resulting in $\boldsymbol{s}=\left[\boldsymbol{s}_{\text{o},1},\boldsymbol{s}_{\text{o},3},\boldsymbol{s}_{\text{o},4},\boldsymbol{s}_{\text{o},6}\right]$. Second, the tagged signal $\boldsymbol{s}_\text{t}$ is obtained by insertion of the frozen tag. Specifically, after the FTG module outputs a frozen tag $\boldsymbol{t}_\text{e}$ with length $N_\text{e}=8$, Alice replaces the raw message $\boldsymbol{s}_\text{o}$ with the frozen tags at the positions specified by set $\mathcal{A}$ to obtain the tagged signal $\boldsymbol{s}_\text{t}=\left[\boldsymbol{t}_{\text{e},1},\boldsymbol{s}_{\text{o},2},\boldsymbol{t}_{\text{e},2},\boldsymbol{t}_{\text{e},3},\boldsymbol{s}_{\text{o},5},\cdots,\boldsymbol{t}_{\text{e},8}\right]$.

	At the receiver, after the Hash Function module determines the index set $\hat{\mathcal{A}}$ of the frozen tags, the SIE module extracts the noisy observation of the frozen tags from the received signal, i.e., $\hat{\boldsymbol{t}}_\text{e}=\left[\boldsymbol{y}_1,\boldsymbol{y}_3,\boldsymbol{y}_4,\boldsymbol{y}_6,\cdots,\boldsymbol{y}_N\right]$. Additionally, the SIE module computes the estimated anchor information $\hat{\boldsymbol{s}}=\left[\hat{\boldsymbol{s}}_{\text{o},1},\hat{\boldsymbol{s}}_{\text{o},3},\hat{\boldsymbol{s}}_{\text{o},4},\hat{\boldsymbol{s}}_{\text{o},6}\right]$ using the first $K_\text{e}$ indices of the set $\hat{\mathcal{A}}$. 
	
	%Since the frozen tags are sparsely embedded in the message, the message bits flipped by these frozen tags can be corrected in the decoding of the message, thereby enhancing the compatibility of the proposed framework. Moreover, 
	Message decoding is compromised when a portion of this message is replaced with frozen tags. Thus, we analyze the impact of tag replacement on the BER of the message in Section \ref{ComAna}, providing guidance for determining the suitable tag insertion ratio.

\vspace*{-0.2cm}	
	\subsection{Paired Frozen Tag Generation and Thaw Tag Reconciliation Modules}
	\vspace*{-0.2cm}
	The paired FTG and TTR modules aim to enhance the robustness of the proposed framework against the user interference and eavesdropping attacks.\footnote{There are multiple channel coding schemes available for the authentication framework proposed in this paper\cite{LDPC,Turbo}. Without loss of generality, polar code is adopted here due to its excellent performance\cite{Polar}.} The FTG module generates frozen tags by placing the raw tag into frozen bits of the polar code and anchor information into the information bits, whereas the TTR module leverages the raw tag as side information to estimate the anchor information for authentication. In this subsection, we continue with the previous example in Section \ref{PSIEMs} to elaborate on these two modules as illustrated in Figure \ref{PolarModule}.
	
	At the transmitter, the FTG module utilizes the anchor information $\boldsymbol{s}$ extracted from the message $\boldsymbol{s}_\text{o}$, and the raw tag $\boldsymbol{t}$ to generate the frozen tags. Specifically, the anchor information $\boldsymbol{s}$ is assigned to the index set of the information bits $\mathcal{I}$, whereas the anchor information is placed at the index set of frozen bits $\mathcal{F}$, i.e., the input vector of encoder $\boldsymbol{u}$ is assembled as
	%\vspace*{-0.2cm}
	\begin{align}\label{III}
		\boldsymbol{u}_\mathcal{I}=\boldsymbol{s},
	\end{align} 
	and
	\begin{align}\label{FFF}
		\boldsymbol{u}_\mathcal{F}=\boldsymbol{t}.
	\end{align}
	For example, for a polar code with code length $N_\text{e}=8$ and code rate $K_\text{e}/N_\text{e}=0.5$, assume that $\mathcal{I} = \left[1, 3, 5, 7\right]$ and $\mathcal{F} = \left[2, 4, 6, 8\right]$. Then, we have the input vector $\boldsymbol{u}=\left[\boldsymbol{s}_1,\boldsymbol{t}_1,\boldsymbol{s}_2,\boldsymbol{t}_2,\boldsymbol{s}_3,\boldsymbol{t}_3,\boldsymbol{s}_4,\boldsymbol{t}_4\right]$. Thus, the frozen tag is given by 
	\vspace*{-0.2cm}
	\begin{align}\label{Gn}
		\boldsymbol{t}_\text{e}=\boldsymbol{u}\mathbf{G},
	\end{align} 
	where $\mathbf{G}$ is the generator matrix of polar code. In the constructed frozen tags, the frozen bits populated with raw tags can facilitate the decoding of the anchor information and resist user interference. Additionally, the frozen bits protect the raw tags from eavesdropping due to the encoding transformation.

	At the receiver, the TTR module recovers the anchor information $\boldsymbol{s}$ from the noisy observation of the frozen tag with the estimated raw tag. Specifically, Bob feeds the noisy codeword $\hat{\boldsymbol{t}}_\text{e}$ into a polar decoder whose a-priori frozen bit distribution is dictated by the raw tag $\boldsymbol{t}$ rather than by the typically all-zero vector. Thus, the decoded reconciled message can be expressed as
%	\vspace*{-0.2cm}
	\begin{align}
		\hat{\boldsymbol{s}}_{\text{Re}} = \text{Dec}\left(\hat{\boldsymbol{t}}_\text{e},\boldsymbol{t}\right),
	\end{align}
	where $\text{Dec}\left(\cdot\right)$ is the decoder of the polar code. The log-likelihood ratios (LLRs) of the frozen indices $\mathcal{F}$ are biased toward the corresponding bits of $\boldsymbol{t}$, whereas the LLRs of the information indices $\mathcal{I}$ are processed in the usual way, i.e.
%	\vspace*{-0.2cm}
	\begin{align}
		{{\boldsymbol{\hat u}}_i} = \left\{ {\begin{array}{*{20}{c}}
				{{{\boldsymbol{t}}_j},}&{i \in {{\mathcal{F}}}}\\
				{{\boldsymbol{\hat u}_{{\text{dec,}}i}},}&{i \in {\mathcal{I}}}
		\end{array}}\right.,
	\end{align}
	where the index $j$ satisfies  $\mathcal{F}_j=i$. ${{\boldsymbol{\hat u}_{{\text{dec,}}i}}}$ denotes the decision outcome of the information bit, i.e. 
	\vspace*{-0.2cm}
	\begin{align}
		{\boldsymbol{\hat u}_{{\text{dec,}}i}} = \left\{ {\begin{array}{*{20}{c}}
				0,&{{\boldsymbol{L}}_N^i\left( {{\boldsymbol{y}},{\boldsymbol{\hat{u}}}_{[i - 1]}} \right) \ge 1}\\
				1,&{{\boldsymbol{L}}_N^i\left( {{\boldsymbol{y}},{\boldsymbol{\hat{u}}}_{[i - 1]}} \right) < 1}
		\end{array}} \right.,	
	\end{align}
	where ${\boldsymbol{L}}_N^i\left( {{\boldsymbol{y}},{\boldsymbol{\hat{u}}}_{[i - 1]}} \right)$ is the LLR for the $i$-th bit of $N$, derived from the received signal $\boldsymbol{y}$ and the estimated bits ${\boldsymbol{\hat{u}}}_{[i - 1]}$.
	
	On the one hand, the proposed scheme withstands user interference by the encoding of the frozen tags. On the other hand, since the raw tags are embedded in the frozen bits, it is difficult for the eavesdropper to recover the raw tags, as will be analyzed in Section \ref{SecAna}.
	
	\begin{remark}
		In contrast to typical PLA schemes, which directly superimpose or insert the raw tags into the message\cite{PLA2008,CRH,XiePSA}, the proposed framework generates frozen tags using both the raw tags and the anchor information. In terms of robustness, the raw tags act as the side information to facilitate the decoding of the anchor information, thereby improving robustness. From a security perspective, the proposed framework offers two advantages. On one hand, since the anchor information is part of the message, the frozen tags are message-dependent.  This prevents an eavesdropper from launching a spoofing attack using the intercepted frozen tags, as these tags are not universally applicable. On the other hand,  since the frozen bits are of low reliability and difficult to decode, it is difficult for an eavesdropper to intercept the raw tags. Regarding system overhead, unlike typical schemes exhibiting linear complexity, our framework incurs additional computational cost primarily due to the processing of frozen tags.  Nevertheless, the overhead remains marginal due to the low quasi-linear complexity of polar codes, i.e., $\mathcal{O}(LN_\text{e}\text{log}_2N_\text{e})$\cite{Polar}, where $L$ is the length of the decoding list.
	\end{remark}

\vspace*{-0.2cm}
\section{Performance Analysis}\label{4}
\vspace*{-0.3cm}

In this section, we analyze the proposed authentication framework in terms of robustness, security, and compatibility. Specifically, regarding robustness, we provide a union bound on the detection probability. In terms of security, both Eve's eavesdropping and spoofing attacks are considered. For compatibility, we provide the upper bound of the BER based on the Bhattacharyya parameters.
\vspace*{-0.4cm}
\subsection{Robustness Analysis}
\vspace*{-0.2cm}
The robustness of the PLA scheme indicates the ability of Bob to distinguish legitimate signals from noisy and interfered observations. In the proposed framework, the FTG and TTR modules are designed to withstand the interference and perform authentication. Thus, the robustness of the authentication depends on the performance of these two modules, i.e., the error performance of the polar code.

In the TTR module, the reconciled message is obtained by hard decision decoding. Thus, a successful authentication is achieved if and only if the reconciled message $\hat{\boldsymbol{s}}_{\text{Re}}\in \mathbb{F}_2^{1\times N_\text{e}}$ matches the estimated message ${\boldsymbol{\hat s}}\in \mathbb{F}_2^{1\times N_\text{e}}$ exactly. (\ref{TS}) can then be rewritten as
\vspace*{-0.2cm}
\begin{align}
	\delta  = \sum\limits_{i = 1}^{{K_{\text{e}}}} {\left( {1 - |{{{\boldsymbol{\hat s}}}_{{\text{Re,}}i}} - {{{\boldsymbol{\hat s}}}_i}|} \right)} \left\{ {\begin{array}{*{20}{c}}
			{ < {K_{\text{e}}},}&{{{\text{H}}_0}}\\
			{ = {K_{\text{e}}},}&{{{\text{H}}_1}}
	\end{array}} \right.,
\end{align}
where the threshold is set to $K_\text{e}$. This fixed threshold is consistent with the properties of polar codes. Specifically, in polar codes, channel polarization splits the original channel into perfect sub-channels with capacity approaching 1 and noisy sub-channels with capacity approaching 0. We place the anchor information in the good sub-channels, i.e., the set $\mathcal{I}$, ensuring reliable transmission of anchor information. Thus, the robustness of the proposed framework depends on the good sub-channels. Due to the difficulty of conducting a theoretical analysis of the channel coding in the proposed framework, we provide a Gaussian approximation-based union bound for the detection probability, i.e.\cite{PDUB}
%\vspace*{-0.2cm}
\begin{align}\label{PDLB}
	{P_{\text{D,UB}}} =1- \mathop \sum \limits_{i \in {\mathcal{I}}}   P\left( {{C_i}} \right) ,
\end{align}
where $C_i\triangleq \left\{\hat{u}_i\neq u_i|\hat{u}_{[i - 1]}= u_{[i - 1]}\right\}$ denotes the event that $\hat{u}_i\neq u_i$ under the condition $\hat{u}_{[i - 1]}= u_{[i - 1]}$ when the SC decoder is adopted. $P\left( {{C_i}} \right)=Q\left(\sqrt{e_i/2}\right)$ is the error probability of the $i$-th bit channel. $e_i$ denotes the reliability of the bit channel, determined through iterative Gaussian approximation \cite{GA} with an initial value of $2/\sigma^2$. %Though generally loose due to ignored error correlation,
Eq. (\ref{PDLB}) offers a tractable, trend-preserving estimation of the detection probability and parameter setting. 
%Note that, when the successive cancellation list (SCL) algorithm\cite{SCL}, which degrades to the successive cancellation (SC) algorithm\cite{Polar} when the list length $L=1$, is employed, the detection probability of the proposed framework increases with the list length. Moreover, 
It can be observed from (\ref{PDLB}) that the authentication performance of the proposed scheme can be improved by flexibly reducing the code rate of the frozen tag.
\vspace*{-0.2cm}
\subsection{Security Analysis}\label{SecAna}
\vspace*{-0.2cm}
%Owing to the open nature of the physical channel, Eve can launch two types of attacks: eavesdropping on the message transmitted by Alice, i.e., eavesdropping attack, and sending forged messages to Bob, i.e., spoofing attack. Thus, 
In this subsection, first, we analyze Eve's capability to infer the secret information from the received messages. Specifically, since Alice's broadcast signal contains inserted frozen tags, Eve may attempt to infer the authentication information. We evaluate the proposed scheme in terms of the resilience against eavesdropping attack. Second, the analysis examines the scheme's defense against spoofing attacks, where Eve forges an authentication signal to deceive Bob despite having no knowledge of the secret key. 
%A critical distinction between Eve and Alice is that Alice possesses full knowledge of the key, whereas Eve has no prior knowledge of it. This limits his probability of successfully launching a spoofing attack.

Consistent with prior work \cite{PLA2008,ATBS,PhaPLA}, the following assumptions are made. First, it is assumed that Eve has perfect knowledge of both the Eve-Alice and Eve-Bob channels. Second, Eve can perfectly eliminate interference from other users. Third, Eve can correctly decode the messages. Note that these assumptions are made in favor of Eve, thereby constituting a worst-case scenario for our security evaluation.
\vspace*{-0.4cm}
\subsubsection{Eavesdropping Attack}\label{EAV}
For the proposed framework, Eve encounters two challenges when attempting an eavesdropping attack. First, without knowledge of the secret key, it is difficult for Eve to determine the exact positions of the frozen tags, i.e., the set $\mathcal{A}$, leading to a Position Confusion Challenge. Second, Eve cannot correctly infer the frozen bits from the estimated frozen tags, resulting in a Tag Confusion Challenge.

\begin{figure}[!t]
	\setlength{\abovecaptionskip}{0pt}
	\centering
	\captionsetup{font={scriptsize}}
	\includegraphics [width=6.2in]{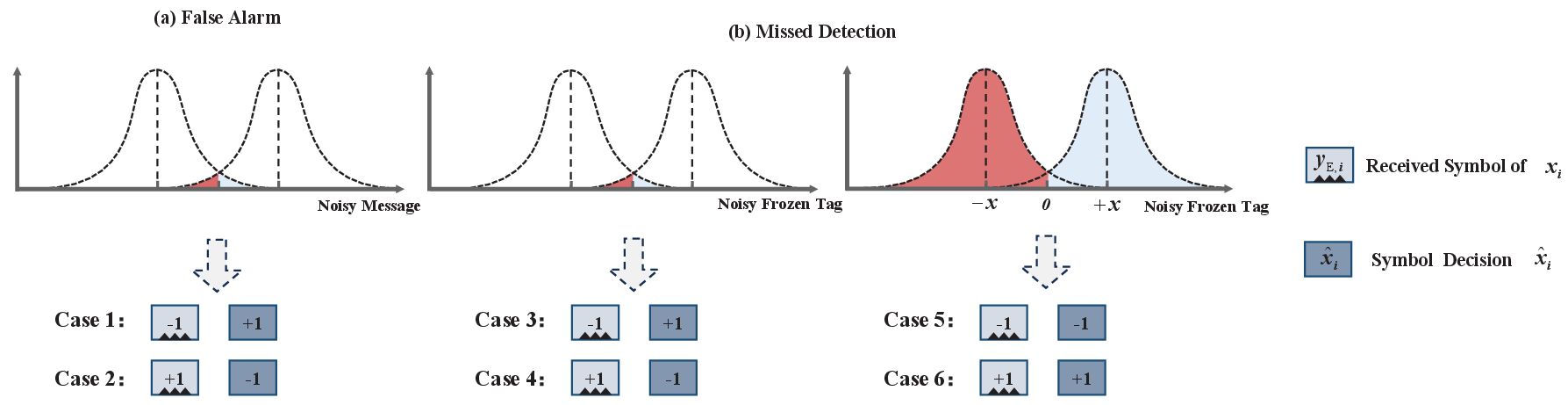}
	\caption{Analysis of Position Confusion under Eavesdropping Attacks, where (a) False alarm, i.e., position of the message is misidentified as that of tag, and (b) Missed detection, i.e., position of the tag is misidentified as that of the message, are included.}
	\label{Eavesdropping}
	\vspace*{-0.3cm}
\end{figure}

\textbf{Position Confusion Challenge:} 
Without loss of generality, the received signal at Eve can be expressed as
\vspace*{-0.2cm}
\begin{align}
	{{\boldsymbol{y}}_{\text{E}}} = {{\boldsymbol{x}}} + {{\boldsymbol{w}}_{\text{E}}},
\end{align}
where ${{\boldsymbol{w}}_{\text{E}}}\sim\mathcal{CN}\left(\boldsymbol{0},\sigma_\text{E}^2\mathbf{I}_N\right)$ is the normalized receiver noise with the covariance $\sigma_\text{E}^2\mathbf{I}_N$ and the signal-to-noise ratio (SNR) is $1/\sigma_\text{E}^2$. As in \cite{XiePSA}, Eve attempts to estimate the positions of the frozen tags by comparing the received signal $\boldsymbol{y}$ with the estimated message $\boldsymbol{s}_\text{o}$, i.e., by performing a bit-wise decision. However, as illustrated in Figure \ref{Eavesdropping}, Eve may experience two types of errors during this process: the false alarm probability, i.e., a message bit is incorrectly identified as a frozen tag bit, and the miss detection probability, i.e., a frozen tag bit is misidentified as a message bit. Specifically, for the false alarm probability, the polarity of a received message symbol may be flipped due to receiver noise as shown in Figure \ref{Eavesdropping}(a). Thus, the false alarm probability is defined as the probability that the polarity of the received symbol differs from that of the corresponding message, i.e.
\vspace*{-0.2cm}
\begin{align}
	{P_{{\text{FA}}|i \in {{\mathcal{A}}^{\text{c}}}}} =& P\left( {{{\boldsymbol{x}}_{i}} =  - 1} \right)\times\underbrace {P\left( {{{\boldsymbol{y}}_{\text{E},i}} > 0|{{\boldsymbol{x}}_{i}} =  - 1,i \in {{\mathcal{A}}^{\text{c}}}} \right)}_{{\text{Case 1}}}+ P\left( {{{\boldsymbol{x}}_{i}} =  + 1} \right)\times\underbrace {P\left( {{{\boldsymbol{y}}_{\text{E},i}} < 0|{{\boldsymbol{x}}_{i}} =  + 1,i \in {{\mathcal{A}}^{\text{c}}}} \right)}_{{\text{Case 2}}}\notag\\
	\overset{\text{A}}{=}& P\left( {{{\boldsymbol{y}}_{\text{E},i}} > 0| {{\boldsymbol{x}}_{i}}=  - 1,i \in {{\mathcal{A}}^{\text{c}}}}\right)= Q\left(\frac{1}{\sigma_\text{E}}\right),
\end{align}
where $\overset{\text{A}}{=}$ holds because $P\left({{\boldsymbol{x}}_{i}}=+1\right)=P\left({{\boldsymbol{x}}_{i}} = -1\right)= 0.5$. For the miss detection probability, as illustrated in Figure \ref{Eavesdropping}(b), two cases should be considered depending on whether the polarity of a frozen tag matches that of the raw message bit. When the polarities differ and the polarity of the frozen tag is flipped by  receiver noise, the polarity of the received symbol becomes identical to that of the raw message. In this case, Eve mistakenly decides that the position corresponds to a message bit, i.e.
\vspace*{-0.2cm}
\begin{align}
	{P_{{\text{MD,1}}|i \in {\mathcal{A}}}} =& P\left({{\boldsymbol{x}}_{i}} =-1\right)\times\underbrace {P\left( {{{\boldsymbol{y}}_{\text{E},i}} > 0|{{\boldsymbol{x}}_{i}} = - 1,i \in {\mathcal{A}}} \right)}_{{\text{Case 3}}}+ P\left( {{{\boldsymbol{x}}_{i}} = + 1} \right)\times\underbrace {P\left( {{{\boldsymbol{y}}_{\text{E},i}} < 0|{{\boldsymbol{x}}_{i}} =  + 1,i \in {\mathcal{A}}} \right)}_{{\text{Case 4}}}\notag\\
	=&Q\left(\frac{1}{\sigma_\text{E}}\right).
\end{align}
In contrast, when the polarity of the frozen tag is identical to that of the raw message and the receiver noise does not flip the frozen tag bit, the polarity of the received symbol remains the same as that of the raw message. Thus, a miss detection also occurs, i.e.
\vspace*{-0.2cm}
\begin{align}
	{P_{{\text{MD,2}}|i \in {\mathcal{A}}}} =& P\left( {{{\boldsymbol{x}}_{i}} =  - 1} \right)\times\underbrace {P\left( {{{\boldsymbol{y}}_{\text{E},i}} < 0|{{\boldsymbol{x}}_{i}} =  - 1,i \in {\mathcal{A}}} \right)}_{{\text{Case 5}}} + P\left( {{{\boldsymbol{x}}_{i}} =  + 1} \right)\times\underbrace {P\left( {{{\boldsymbol{y}}_{\text{E},i}} > 0|{{\boldsymbol{x}}_{i}} =  + 1,i \in {\mathcal{A}}} \right)}_{{\text{Case 6}}}\notag\\
	=&Q\left(\frac{-1}{\sigma_\text{E}}\right).
\end{align}
Thus, for any given bit, the average error probability at Eve can be expressed as
\vspace*{-0.2cm}
\begin{align}\label{perr}
	{P_{{\text{err}}}} =& P(i \in {{\cal A}^{\text{c}}})\times{P_{{\text{FA}}|i \in {{\mathcal{A}}^{\text{c}}}}} + P(i \in {\mathcal{A}})\times\left( {{P_ {\neq} }{P_{{\text{MD,1}}|i \in {\mathcal{A}}}} + {P_{=}}{P_{{\text{MD,2}}|i \in {\mathcal{A}}}}} \right),
\end{align}
where $P(i \in {\mathcal{A}}) = {{N_\text{e}}}/N$ denotes the probability that an arbitrary position corresponds to a frozen tag bit, while $P(i \in {{\mathcal{A}}^{\text{c}}}) = 1 - P(i \in {\mathcal{A}})$ denotes the probability that it corresponds to a message bit. In addition, ${P_{=}}$ represents the probability that the frozen tag has the same polarity as the corresponding message bit, whereas ${P_{\neq} }$ denotes the probability that their polarities differ. From (\ref{perr}), it is easy to obtain the $P_\text{err}$ in the high SNR regime, i.e.
\vspace*{-0.2cm}
\begin{align}
	P_\text{err,asy}=P_{\neq}\times\frac{N_\text{e}}{N}.
\end{align}
Moreover, the probability that each bit in the received signal is correctly classified by Eve is given by
\vspace*{-0.2cm}
\begin{align}
	P_{\text{PCC}}=\left(1-P_\text{err}\right)^N.
\end{align}
For example, when we set SNR $=2$ dB, $N_\text{e}=128$, and $N=256$, we have $P_{\text{err}}=30.2\%$ and $P_\text{PCC}\approx 10^{-40}$. Thus, it can be seen that it is extremely difficult for Eve to overcome the Position Confusion Challenge.

\textbf{Tag Confusion Challenge:} Furthermore, even if Eve is extremely fortunate and correctly guesses the position of the frozen tag $\mathcal{A}$, it would still be challenging for him to estimate the raw tag without the secret key. Specifically, the noisy frozen tag can be expressed as
\vspace*{-0.2cm}
\begin{align}
	{{\boldsymbol{y}}_{{\text{E,}}{\cal A}}} ={\cal M}\left( {{{\boldsymbol{s}}_{{\text{t,}}{\cal A}}}} \right) + {{\boldsymbol{w}}_{{\text{E,}}{\cal A}}},
\end{align}
where $\mathcal{M}\left(\boldsymbol{x}\right)$ denotes the modulation mapping. For BPSK, the mapping is given by ${\cal M}\left( {\boldsymbol{x}} \right) = 1 - 2{\boldsymbol{x}}$. To facilitate the analysis, we map the received signal from the modulation domain to the coding domain, i.e. 
\begin{align}
	{{\boldsymbol{\tilde y}}_{{\text{E,}}{\cal A}}} = {{\boldsymbol{s}}}{{\mathbf{G}}_{\mathcal{I}}} + {{\boldsymbol{t}}}{{\mathbf{G}}_{\cal F}} + {{\boldsymbol{\tilde w}}_{{\text{E,}}{\cal A}}},
\end{align}
where ${{\boldsymbol{\tilde w}}_{{\text{E,}}{\cal A}}} =  - \frac{{{{\boldsymbol{w}}_{{\text{E,}}{\cal A}}}}}{2}
\sim {\cal C}{\cal N}\left( {{\boldsymbol{0}},\frac{{\sigma _{\text{E}}^2}}{4}{{\mathbf{I}}_{{N_{\text{e}}}}}} \right)$ and ${{\bf{G}}_{\cal D}} \in \mathbb{F}_2^{|{\cal D}| \times  {N_{\text{e}}}}$, ${\cal D} \in \left\{ {{\mathcal{I}},{\mathcal{F}}} \right\}$, is formed by the rows of the generator matrix $\bf{G}$ that are indexed by the set $\mathcal{D}$. Thus, we have 
\begin{align}
	{{\boldsymbol{\tilde y}}_{{\text{E,}}{\cal F}}} ={{\boldsymbol{t}}}{{\bf{G}}_{\cal F}} + {{\boldsymbol{\tilde w}}_{{\text{E,}}{\cal A}}}.
\end{align}
Since the raw tag $\boldsymbol{t}$ is placed in the frozen bits, Eve faces difficulty in estimating the raw tag. Thus, we analyze the Tag Confusion Challenge by examining the SNR of the raw tag $\boldsymbol{t}$. Specifically, since ${{\bf{G}}_{\cal F}}$ is row full-rank, guaranteed by the structure of the polar code, there exists a matrix ${{\bf{M}}_{\text{R}}} = {\bf{G}}_{\cal F}^\mathsf{T}{\left( {{{\bf{G}}_{\cal F}}{\bf{G}}_{\cal F}^\mathsf{T}} \right)^{ - 1}} \in \mathbb{F}_2^{{N_{\text{e}}} \times  {\left(N_\text{e}- {K_{\text{e}}}\right) } }$  such that ${{\bf{G}}_{\cal F}}{{\bf{M}}_{\text{R}}} = {\mathbf{I}_{ N_\text{e}- {K_{\text{e}}} }}$. Thus, the raw tag can be estimated by 
\vspace*{-0.2cm}
\begin{align}
	{\hat{ \boldsymbol{t}} }= {{\boldsymbol{t}}} + {\boldsymbol{\tilde w}},
\end{align}
where ${\boldsymbol{\tilde w}} = {{\boldsymbol{\tilde w}}_{{\text{E,}}{\cal A}}}{{\bf{M}}_{\text{R}}}$ is the accumulated noise. Since ${{\boldsymbol{\tilde w}}_{{\text{E,}}{\cal A}}}$  follows a CSCG distribution and ${\boldsymbol{\tilde w}}$  is a linear transformation of ${{\boldsymbol{\tilde w}}_{{\text{E,}}{\cal A}}}$, we have ${\boldsymbol{\tilde w}}\sim {\cal C}{\cal N}\left( {{\boldsymbol{0}},{\Sigma _{{\boldsymbol{\tilde w}}}}} \right)$, where
\begin{align}\label{Sigmaw}
	{\Sigma _{{\boldsymbol{\tilde w}}}} = \frac{{\sigma _{\text{E}}^2}}{4}{\left( {{{\bf{G}}_{\cal F}}{\bf{G}}_{\cal F}^\mathsf{T}} \right)^{ - 1}}.
\end{align}
We can observe from (\ref{Sigmaw}) that when Eve attempts to estimate the raw tag, the effective SNR depends on the set of frozen bit $\mathcal{F}$ and the code length $N_\text{e}$. Specifically, the estimation of the raw tag essentially projects the $N_\text{e}$-dimensional receiver noise ${{\boldsymbol{\tilde w}}_{{\text{E,}}{\cal A}}}$ onto the $(N_\text{e}-K_\text{e})$-dimensional raw tag subspace via a pseudo-inverse mapping ${{\bf{M}}_{\text{R}}}$. This linear transformation causes noise accumulation, thereby significantly degrading the SNR.

\begin{remark}
	It can be observed that  the proposed scheme can effectively defend against eavesdropping attacks. Specifically, due to a high probability of misclassification in (\ref{perr}),  Eve struggles to accurately estimate the position of the frozen tag. Furthermore, due to significant noise accumulation as in (\ref{Sigmaw}), it is also difficult for Eve to estimate the raw tag correctly.
\end{remark}

\vspace*{-0.3cm}
\subsubsection{Spoofing Attack}
In the spoofing attack, Eve faces difficulty in determining the set of positions for the frozen tag and in generating codewords with the correct raw tag as frozen bits without the knowledge of the secret key.

Specifically, Eve feeds the anchor information and the raw tag into the FTG module as in (\ref{Gn}). However, since Eve is unaware of the secret key, he can only populate the frozen bits with a random raw tag, i.e.
\begin{align}
	{\boldsymbol{u}}_{\mathcal{F}} = {{\boldsymbol{t}}_{\text{E}}}.
\end{align}
Furthermore, as indicated by Section \ref{EAV}, Eve faces difficulty in estimating the exact positions of the frozen tag $\mathcal{A}$. Thus, the position of the frozen tag for Eve $\mathcal{A}_\text{E}$ is also random, i.e.
\begin{align}
	{{\cal A}_{\text{E}}} = {\text{Ge}}{{\text{n}}_{{\text{Pos}}}}\left( {{{\boldsymbol{s}}_{{\text{Eo}}}},{{\boldsymbol{k}}_{\text{E}}}} \right),	
\end{align}
where $\boldsymbol{k}_\text{E}$ is the random secret key of Eve. The tagged signal is obtained by inserting frozen tags into the message, i.e.
\vspace*{-0.2cm}
\begin{align}
	\boldsymbol{s}_{\text{E},\mathcal{A}_\text{E}}=\boldsymbol{t}_\text{Ee},
\end{align}
and
\begin{align}
	\boldsymbol{s}_{\text{E},\mathcal{A}_\text{E}^{\text{c}}}=\boldsymbol{s}_{\text{Eo},\mathcal{A}_\text{E}^\text{c}},
\end{align}
where $\boldsymbol{t}_\text{Ee}$ is the frozen tag and $\boldsymbol{s}_\text{Eo}$ is the message. Thus, the frozen tag and the position of tag insertion are both random.

At the receiver, to extract the frozen tags, Bob first estimates the position of the frozen tag $\mathcal{A}_\text{B}$, i.e.
\begin{align}
	{{\cal A}_{\text{B}}} = {\text{Ge}}{{\text{n}}_{{\text{Pos}}}}\left( {{{{\boldsymbol{\hat s}}}_{{\text{Eo}}}},{\boldsymbol{k}}} \right),	
\end{align}
where $\hat{\boldsymbol{s}}_\text{Eo}$ is the estimation for the message of Eve $\boldsymbol{s}_\text{Eo}$. Note that, since $\mathcal{A}_\text{B}$ is generated by the correct raw tag, it is independent of Eve's frozen tag positions $\mathcal{A}_\text{E}$. Thus, the alignment between $\mathcal{A}_\text{B}$ and $\mathcal{A}_\text{E}$ becomes elusive. Due to the difficulty in analyzing the hash function, we characterize the difference between $\mathcal{A}_\text{B}$ and $\mathcal{A}_\text{E}$ by introducing the symmetric difference\cite{SD}, which consists of elements that belong to only one of the two sets, i.e.
\vspace*{-0.2cm}
\begin{align}
	|{{\cal A}_{\text{B}}}\Delta {{\cal A}_{\text{E}}}| \buildrel \Delta \over = |{{\cal A}_{\text{B}}} \cup {{\cal A}_{\text{E}}}| - |{{\cal A}_{\text{B}}} \cap {{\cal A}_{\text{E}}}|= 2\left({N_{\text{e}}} - |{{\cal A}_{\text{B}}} \cap {{\cal A}_{\text{E}}}|\right).
\end{align}
To obtain the expectation of the symmetric difference between $\mathcal{A}_\text{B}$ and $\mathcal{A}_\text{E}$, we define an indicator variable $\boldsymbol{I}_i$, $i\in[N]$, such that 
\vspace*{-0.2cm}
\begin{align}
	{{\boldsymbol{I}}_i} = \left\{ {\begin{array}{*{20}{c}}
			1,&{i \in \left( {{{\cal A}_{\text{B}}} \cap {{\cal A}_{\text{E}}}} \right)}\\
			0,&{{\text{otherwise}}}
	\end{array}} \right.,	
\end{align}
Thus, the expectation of the symmetric difference can be expressed as
\vspace*{-0.2cm}
\begin{align}
	\mathbb{E}\left( {|{{\cal A}_{\text{B}}} \cap {{\cal A}_{\text{E}}}|} \right) = \sum\limits_{i = 1}^N \mathbb{E}{\left( {{{\boldsymbol{I}}_i}} \right)} 
	\overset{\text{A}}{=}  \frac{{N_{\text{e}}^2}}{{{N}}}, 
\end{align}
where $\overset{\text{A}}{=}$ holds since $\mathcal{A}_\text{E}$ and $\mathcal{A}_\text{B}$ are independent of each other. Thus, the expectation of the normalized symmetric difference is given by 
\vspace*{-0.2cm}
\begin{align}
	P_\text{SD}\triangleq\frac{\mathbb{E}	\left( {|{{\cal A}_{\text{B}}}\Delta {{\cal A}_{\text{E}}}|} \right)}{2N_\text{e}} 
	=  {1 - \frac{{{N_{\text{e}}}}}{N}}.
\end{align}
We can observe that the expected ratio of common elements between sets $\mathcal{A}_\text{E}$ and $\mathcal{A}_\text{B}$ decreases as $N$ increases. Thus, the position for Eve's frozen tag becomes increasingly difficult to align with that of Bob.

Furthermore, even if the sets happen to align with each other by chance, it is difficult for him to construct the correct frozen tag due to the lack of the secret key.

\begin{figure}[!t]
	\setlength{\abovecaptionskip}{0pt}
	\centering
	\captionsetup{font={scriptsize}}
	\includegraphics [width=3.5in]{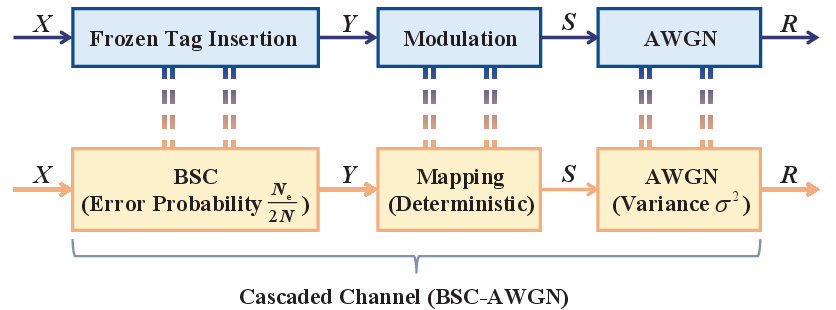}
	\caption{Cascaded channel, where the first one is a BSC channel modeled by frozen tag insertion while the second one is an AWGN channel.}
	\label{Cchannel}
	\vspace*{-0.3cm}
\end{figure}
\vspace*{-0.3cm}
\subsection{Compatibility Analysis}\label{ComAna}
\vspace*{-0.2cm}

Compatibility refers to the property that the authentication scheme does not affect communication performance for an unconscious receiver. In the proposed framework, the frozen tag is randomly inserted into the message using a secret key. To ensure good compatibility, the insertion ratio must be carefully set. Due to the challenge of deriving a closed-form expression for the error performance of an unconscious receiver, we model the tag insertion and wireless transmission processes as a cascaded channel. The compatibility of the scheme is then evaluated numerically through the Bhattacharyya parameter.

Specifically, at the transmitter, the frozen tag is randomly inserted into the message to obtain a tagged signal $\boldsymbol{s}_\text{t}$ with an insertion ratio of $N_\text{e}/N$. This tagged signal is then modulated to produce the transmitted signal. The unconscious receiver treats the noisy observation as a standard polar code and decodes it using SC or SCL decoding. For simplicity, the BPSK modulation and an AWGN channel are assumed here. Thus, we model the tag insertion and wireless transmission as a cascaded channel X$\in\left\{0,1\right\}$$\rightarrow$Y$\in\left\{0,1\right\}$$\rightarrow$S$\in\left\{\pm1\right\}$$\rightarrow$R$\in\mathbb{R}$, as shown in Figure \ref{Cchannel}. The tag insertion is modeled as a  BSC, with a transition probability of
\vspace*{-0.2cm}
\begin{align}
	{P_{{\text{BSC}}}}\left( {Y = y|X = x} \right) = \left\{ {\begin{array}{*{20}{c}}
			{1 - {P_{{\text{BSC}}}},}&{y = x}\\
			{{P_{{\text{BSC}}}},}&{y \ne x}
	\end{array}} \right.,
\end{align}
where ${P_{{\text{BSC}}}} = \frac{{{N_{\text{e}}}}}{{2N}}$ and  the coefficient $\frac{1}{2}$ indicates the probability that the frozen tag differs from the raw message at the corresponding position.  The second channel is characterized by an AWGN channel with a transition probability of
\vspace*{-0.2cm}
\begin{align}
	{P_{\text{N}}}\left( {R = r|S = s} \right) = \frac{1}{{\sqrt {2\pi {\sigma ^2}} }}{\text{exp}}\left( { - \frac{{{{\left( {r - s} \right)}^2}}}{{2{\sigma ^2}}}} \right).
\end{align}
Thus, the transition probability of the cascaded channel can be given by
\vspace*{-0.2cm}
\begin{align}
	P_{\text{C}}\left( {r|x} \right) = \sum\limits_y {{P_{{\text{BSC}}}}\left( {y|x} \right) \times {P_{\text{N}}}\left( {r|s = 1 - 2y} \right)}.
\end{align}
It is easy to obtain that $P_\text{C}\left( {r|x = 0} \right) = P_\text{C}\left( { - r|x = 1} \right)$. Thus, the cascaded channel is symmetric and the Bhattacharyya parameters of each sub-channels can be iteratively computed\cite{Polar}. Here, the initial value of the sub-channel is given by
\vspace*{-0.2cm}
\begin{align}
	Z\left( {W_N^{(1)}} \right)\!\! = \int_{ - \infty }^{ + \infty } {\sqrt {P_{\text{C}}\left( {r|0} \right)P_{\text{C}}\left( {r|1} \right)} } dr = {\text{exp}}\left( {\! - \frac{1}{{2{\sigma ^2}}}}\! \right)\! \cdot\! \mathbb{E}{_r}\!\left(\sqrt{ {a\! +\! 2b\cosh \left( {\frac{{2r}}{{{\sigma ^2}}}} \right)}} \right),
\end{align}
where $a = {\left( {1 - {P_{{\text{BSC}}}}} \right)^2} + P_{{\text{BSC}}}^2$ and $b = \left( {1 - {P_{{\text{BSC}}}}} \right){P_{{\text{BSC}}}}$. $\mathbb{E}_r\left(\cdot\right)$ denotes the expectation over the random variable $r\sim \mathcal{CN}\left(0,\sigma^2\right)$. Especially, when no tag is inserted, we have
\vspace*{-0.2cm}
\begin{align}
	{Z_0}\left( {W_N^{(1)}} \right) = {\text{exp}}\left( { - \frac{1}{{2{\sigma ^2}}}} \right).
\end{align}
Note that, the transmitter uses the Bhattacharyya parameters corresponding to the AWGN channel, rather than those of the cascaded channel, to determine the reliable sub-channels, i.e., $\mathcal{I}$. Thus, the upper bound of the average BER for the unconscious receiver can be expressed as
\begin{align}
	{P_{\text{Ave}}} \le \frac{1}{|{\mathcal {I}}|}\sum\limits_{i \in {\mathcal {I}}} {Z\left( {W_N^{\left( i \right)}} \right)}. 	
\end{align}
In practical applications, the above bounds can provide guidance on the insertion ratio of frozen tags in a single polar codeword. 
%Furthermore, when the length of the frozen tag $N_\text{e}$ is extremely large, the tag can be dispersedly inserted into multiple polar codewords, thereby reducing the insertion ratio in each codeword. This approach can further enhance the compatibility without altering the fundamental principle of the proposed framework.
\vspace*{-0.5cm}

\begin{figure*}[t!]
	\centering
	\begin{minipage}[t]{0.49\linewidth}
		\centering
		\includegraphics[width=2.7in]{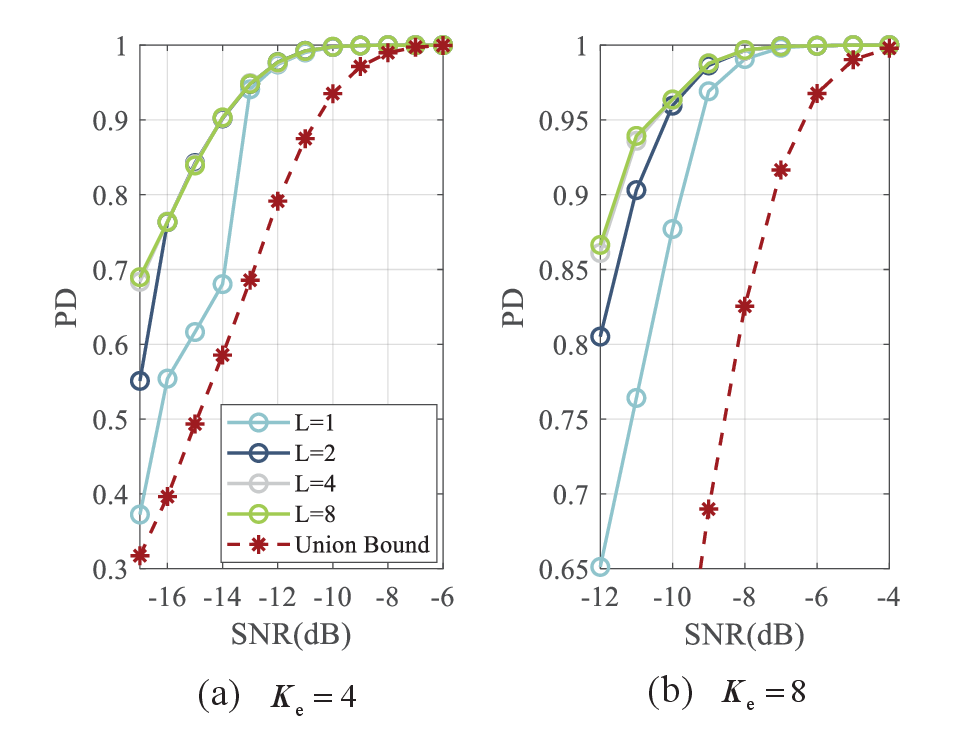}
		\caption{Detection performance versus SNR for different list lengths $L$, with anchor information lengths (a) $K_\text{e} = 4$ and (b) $K_\text{e} = 8$, decoded via SCL decoder ($L = 1$ corresponds to SC decoder).}
		\label{UnionBound}
	\end{minipage}
	\hfill
	\begin{minipage}[t]{0.49\linewidth}
		\centering
		\includegraphics[width=2.8in]{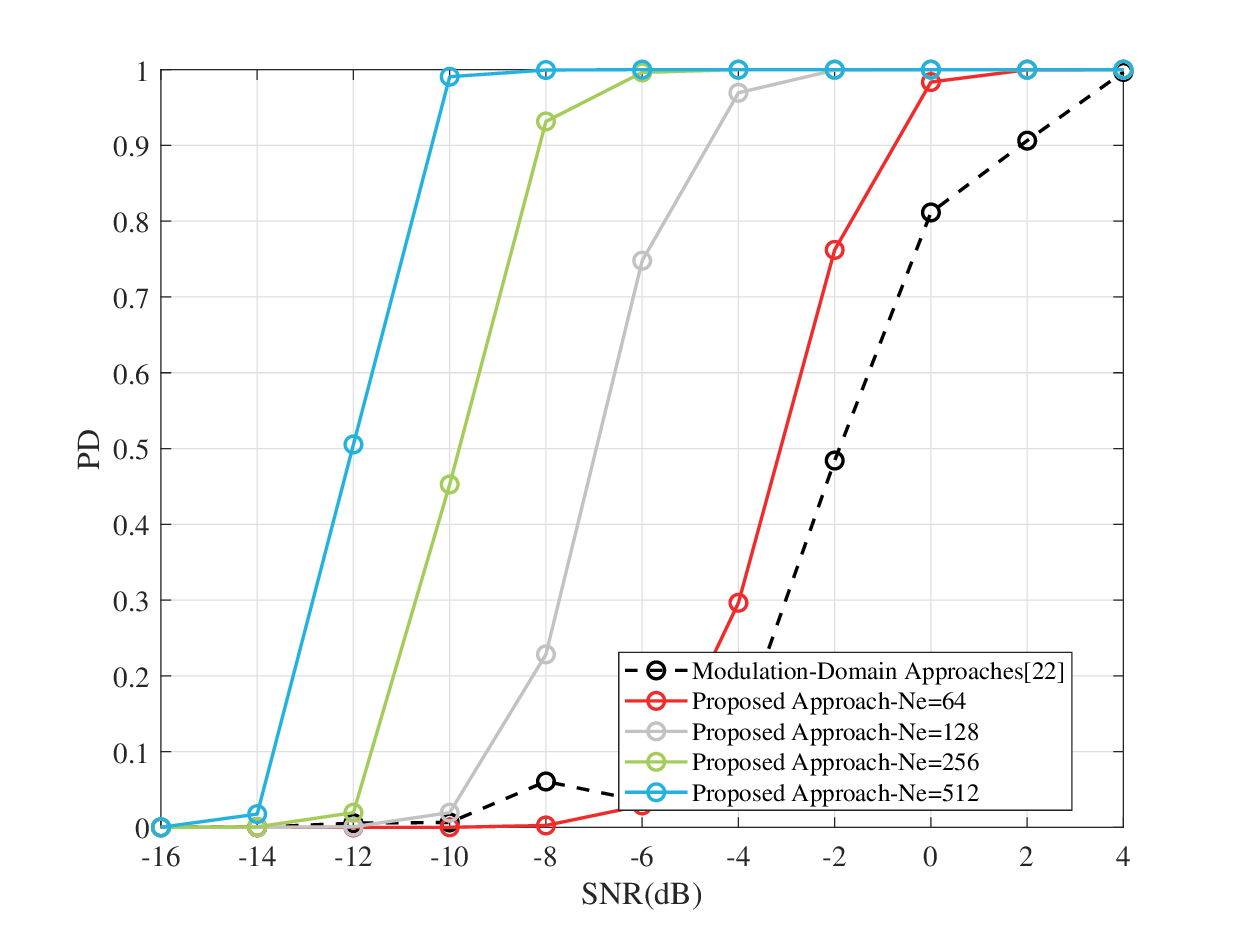}
		\caption{Detection performance of the proposed framework against the uncoded tag-based approach versus SNR for different lengths of frozen tag $N_\text{e}$, where the length of the anchor information $K_\text{e} = 32$.}
		\label{PDFT}
	\end{minipage}
	\vspace*{-0.2cm}
\end{figure*}

\section{Numerical Results}\label{5}
\vspace*{-0.3cm}
In this section, we present the experimental setup and numerical results. Specifically, we first provide the overview of the numerical results and the experimental setup. Second, the performance of the proposed framework regarding the robustness, security, and compatibility  is simulated and analyzed respectively.

\vspace*{-0.2cm}
\subsection{Experimental Setup}
\vspace*{-0.2cm}
In this section, we present the experimental setup. In terms of robustness, we compare the proposed scheme with existing uncoded tag-based benchmark scheme\cite{CRH}, where the raw tags rather than the frozen tags are randomly inserted into the message. Regarding security, we comprehensively simulate the performance of the proposed scheme under eavesdropping and spoofing attacks. In terms of compatibility, we provide the upper bound of the BER at the unconscious receiver, both with and without tag insertion, to guide the setting of insertion ratio in practical applications. Unless otherwise specified, we set the length of the frozen tag  $N_\text{e}=128$ and the length of anchor information  $K_\text{e}=4$. The number of the user interference is $K=8$. The SCL algorithm is adopted, which degenerates to the SC algorithm when the list length is set to 1. Each simulation is performed using $10^5$ independent Monte Carlo experiments.

\vspace*{-0.3cm}
\subsection{Robustness Evaluation}
\begin{figure*}[t!]
	\centering
	\begin{minipage}[t]{0.49\linewidth}
		\centering
		\includegraphics[width=2.35in]{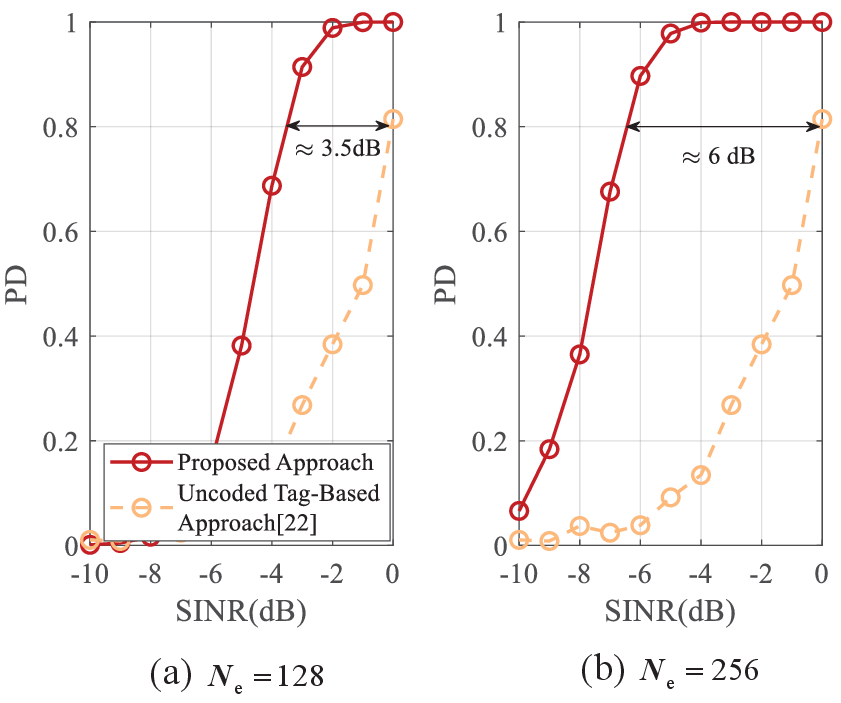}
		\caption{Detection performance of the proposed framework against the uncoded tag-based approach versus SINR with SNR = 0 dB, where (a) $N_\text{e}=128$, and (b) $N_\text{e}=256$.}
		\label{PD_SINR}
	\end{minipage}
	\hfill
	\begin{minipage}[t]{0.49\linewidth}
		\centering
		\includegraphics[width=2.8in]{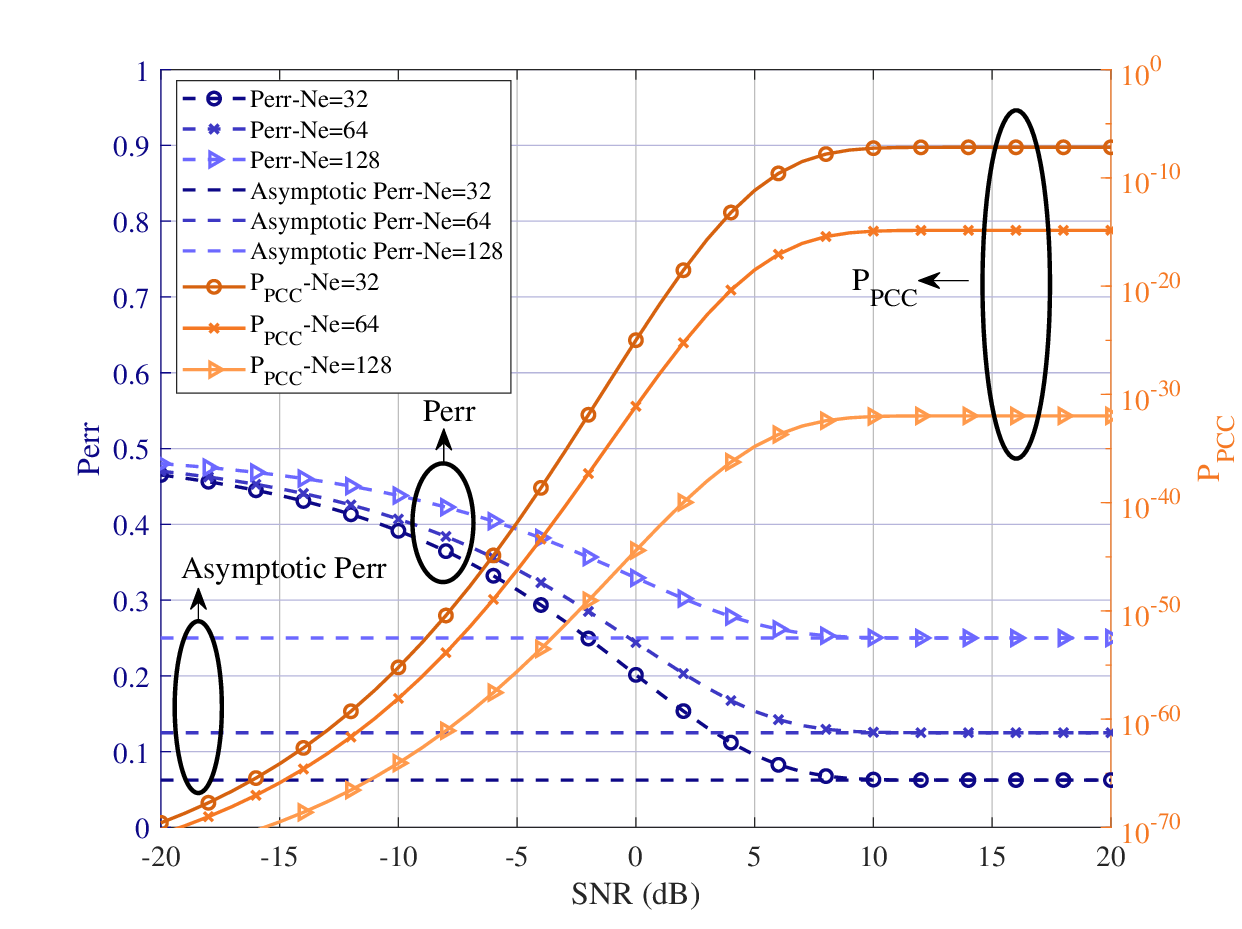}
		\caption{Performance of Eve's estimation for the tag positions in the Position Confusion Challenge versus SNR, where the length of the message $N=256$.}
		\label{PerrAndPpcc}
	\end{minipage}
	%\vspace*{-0.3cm}
\end{figure*}

\vspace*{-0.2cm}
\begin{observation}
	\textit{The proposed framework outperforms uncoded tag-based baselines in detection performance even under the user interference. (cf. Figures \ref{UnionBound}, \ref{PDFT}, and \ref{PD_SINR})}
\end{observation}

\noindent First, we evaluate the authentication performance of the SCL decoder versus SNR for various list lengths L, with the length of the frozen tag $N_\text{e}=128$, and lengths of the anchor information (a) $K_\text{e}=4$ and (b) $K_\text{e}=8$. As depicted in Figure \ref{UnionBound}, the following observation can be made. First, the authentication performance improves as the SNR increases. Second, the authentication performance continuously improves with the increase in the length of the list. This is attributed to the enhanced accuracy of the reconciled message in the TTR module as the length of the list increases. Third, the union bound lies below the simulated curve for $L = 1$ and asymptotically aligns with the Monte Carlo at high SNR. This suggests that the union bound in (\ref{PDLB}) is suitable for quickly evaluating the detection performance of the proposed framework and offering parameter recommendations.

Second, we evaluate the proposed framework against the conventional scheme under different lengths of the frozen tag, with the the length of the anchor information fixed at $K_\text{e}=32$. For fairness in the comparison of robustness, we keep the power of the baseline schemes consistent with that of the proposed framework, resulting in the same detection probability for all baseline schemes.
As shown in Figure \ref{PDFT}, the following phenomena can be observed. First, the detection performance improves  with increasing length of the frozen tags. This is because the additional tags provide extra protection to the anchor information. Second, the proposed framework outperforms the typical scheme, which corroborates the superior performance of the proposed framework.

%\begin{figure}[!t]
	%\centering
	%\captionsetup{font=scriptsize}
	%\setlength{\subfigcapskip}{0pt}
	%\setlength{\abovecaptionskip}{4pt}
	
	%\subfigure[$N_\text{e}=128$]{
	%	\includegraphics[width=0.24\textwidth]{YL_pic/PD_SINR_32_128.eps}
	%	\label{PDSINR1}
	%}
	%\hspace{-20pt}  % ? ???????????? 3pt / 5pt / 0.5em?
	%\subfigure[$N_\text{e}=256$]{
	%	\includegraphics[width=0.24\textwidth]{YL_pic/PD_SINR_32_256.eps}
	%	\label{PDSINR2}
	%}
	
	%\caption{Detection performance of the proposed framework against the uncoded tag-based approach versus SINR %with SNR = 0 dB, where (a) $N_\text{e}=128$, and (b) $N_\text{e}=256$.}
%	\label{PDSINR}
%\end{figure}

Third, we compare the detection performance of the proposed framework with that of the uncoded tag-based scheme under different SINR, given SNR = 0 dB, $K_\text{e}=32$, and $8$ unintended users. Similarly, the tag power of the baseline schemes is kept consistent with the proposed framework. As depicted in Figure \ref{PD_SINR}, the following observation can be made. First, the detection performance of the proposed framework significantly outperforms that of the uncoded tag-based scheme.  Second, the detection performance of the proposed framework improves steadily as the length of the frozen tag  increases. Taking the $80\%$ detection probability as an example, the gap in required SINR between the proposed scheme and the uncoded tag-based scheme is about 3.5 dB when $N_\text{e} = 128$ while this gap increases to 6 dB when $N_\text{e}=256$. This indicates that the proposed framework effectively suppresses user interference through the design of the frozen tag, thereby significantly enhancing the authentication performance. However, typical schemes retain a computational advantage due to the absence of frozen tag processing. Consequently, a focus of our future research is to reduce complexity while enhancing detection performance.

\begin{figure*}[t!]
	\centering
	\begin{minipage}[t]{0.49\linewidth}
		\centering
		\includegraphics[width=2.8in]{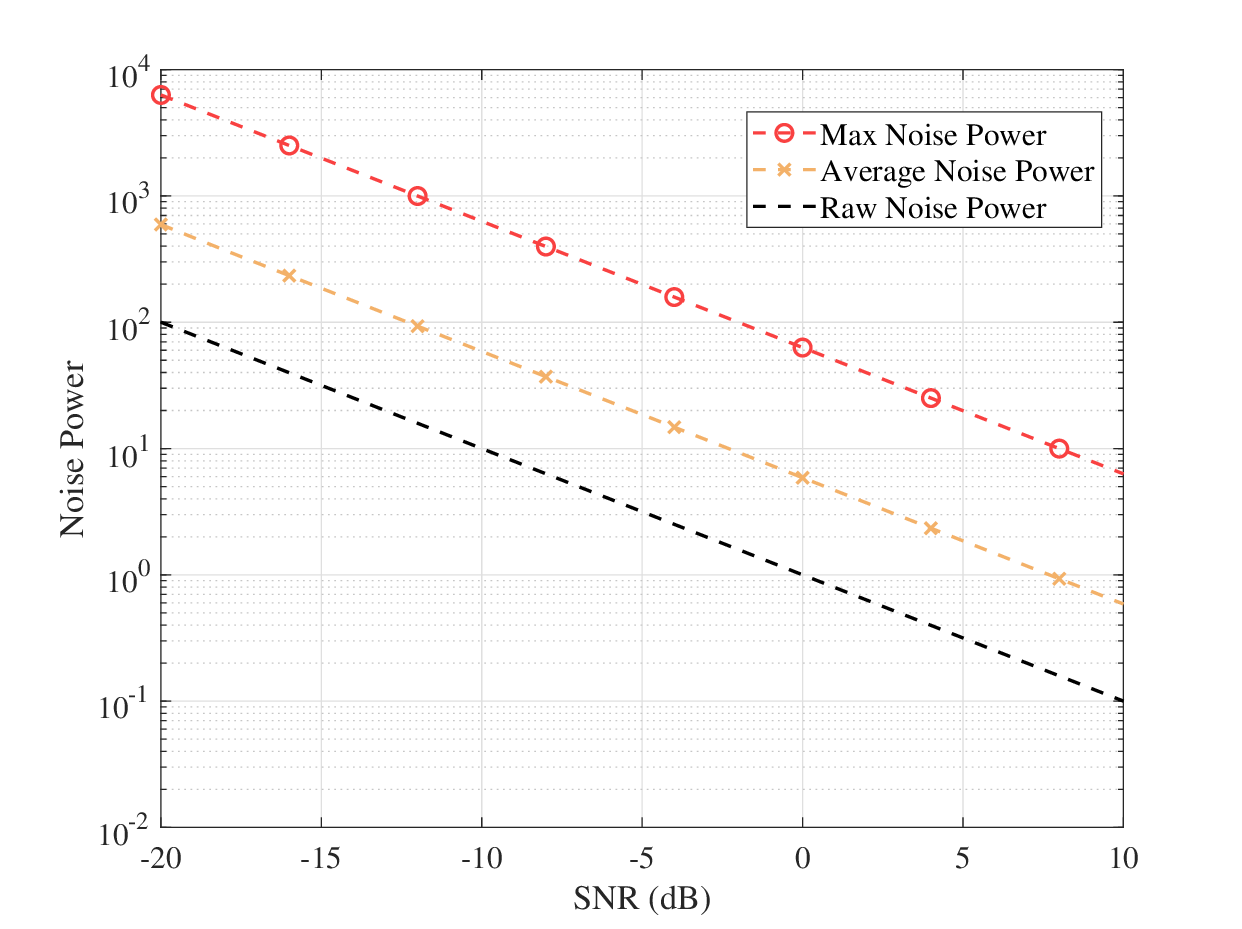}
		\caption{The noise power versus SNR when Eve tries to estimate the raw tag, showing that both the maximum and average noise powers are significantly greater than the receiver noise, i.e., the raw noise power.}
		\label{TagConfusion1}
		\end{minipage}
		\hfill
		\begin{minipage}[t]{0.49\linewidth}
			\centering
			\includegraphics[width=2.8in]{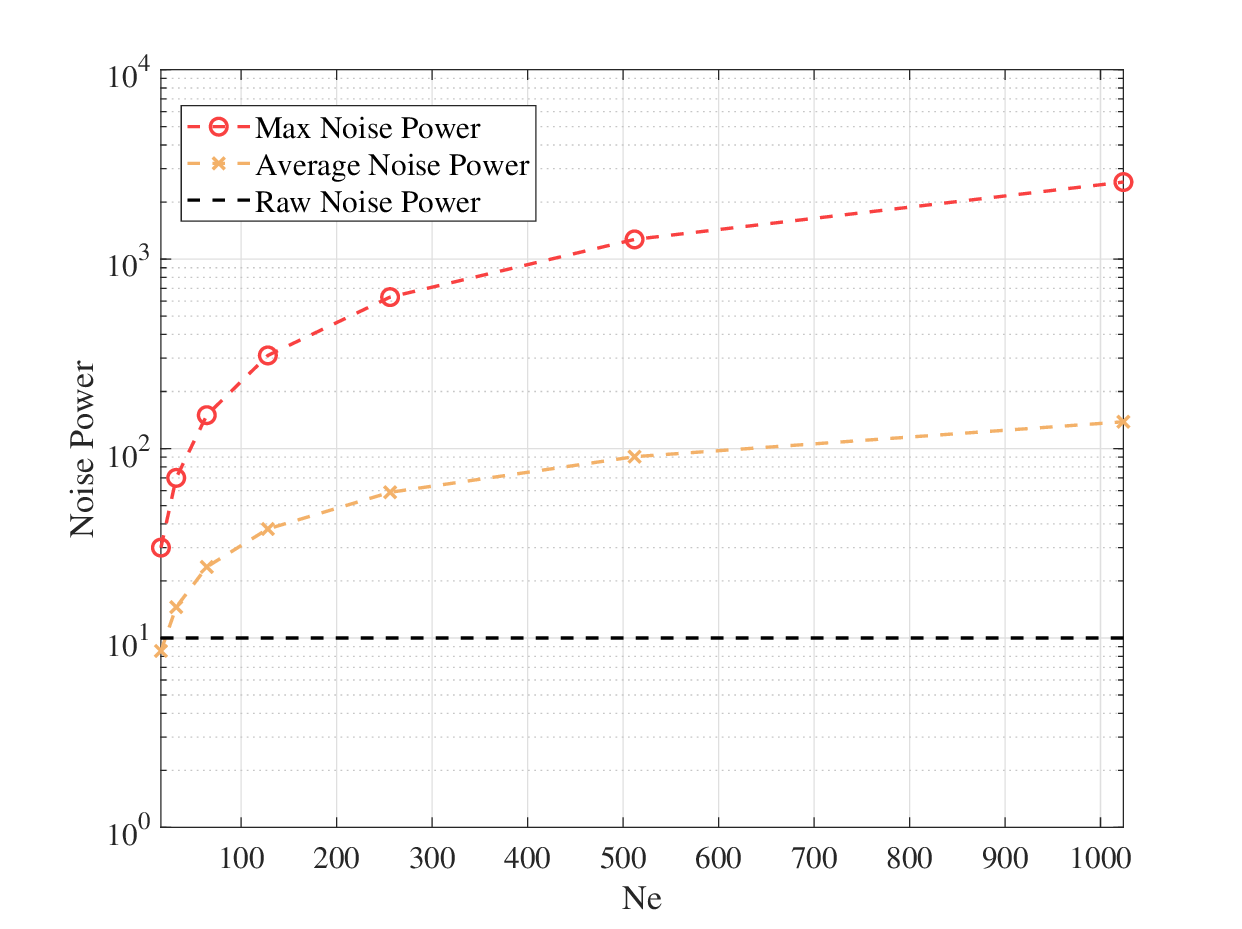}
			\caption{ The noise power versus the length of the frozen tag $N_\text{e}$ when Eve tries to estimate the raw tag, where the maximum and average noise powers are also much greater than the receiver noise.}
			\label{TagConfusion2}
		\end{minipage}
		\vspace*{-0.5cm}
\end{figure*}

\vspace*{-0.3cm}
\subsection{Security Evaluation}
\vspace*{-0.2cm}
\begin{observation}
	\textit{The proposed framework can effectively resist eavesdropping attacks, including Eve's estimation of the tag position and the raw tags. (cf. Figures \ref{PerrAndPpcc}, \ref{TagConfusion1}, and \ref{TagConfusion2})}
\end{observation}
\vspace*{-0.2cm}
\noindent In the position confusion challenge, Eve's average bit error probability and the probability of correctly detecting all frozen tag positions are simulated as shown in Figure \ref{PerrAndPpcc}, where the length of the message is set to $N=256$,  and the lengths of the frozen tags are 32, 64, and 128, respectively. The following observations can be made. First, the average bit error probability $P_\text{err}$ decreases as the SNR increases. This is because the probability of symbol polarity inversion caused by noise diminishes as the power of the noise decreases. Second, Eve's average bit error probability $P_\text{err}$ does not approach zero but rather converges to a given asymptotic error probability in high SNR region. This is due to the fact that when the polarity of the frozen tag matches that of the raw message, it remains difficult for Eve to distinguish the frozen tag at that position even with a high SNR. This provides inherent protection for the security of the proposed scheme. Third, the probability of Eve correctly classifying all positions $P_{\text{PCC}}$ increases with SNR, but even in the high SNR region, the probability remains well below 1\%. Fourth, as the length of the frozen tag increases, the average bit error probability $P_\text{err}$ increases, while the probability of correct classification $P_{\text{PCC}}$ decreases. This suggests that increasing the ratio of frozen tag $N_\text{e}/N$ can reduce the probability of Eve correctly identifying the positions of the frozen tag.

In the tag confusion challenge, we simulate the noise accumulation in Eve's estimation of the raw tag, as shown in Figures \ref{TagConfusion1} and \ref{TagConfusion2}. Specifically, first, the variation of accumulated noise in different SNR is investigated in Figure \ref{TagConfusion1}. It is evident that the accumulated noise decreases as the SNR increases. However, the maximum and average accumulated noise powers remain significantly greater than the raw receiver noise, indicating that Eve always faces substantial noise power when estimating the raw tags. Second, we simulate the variation of noise power with different length of the frozen tag in Figure \ref{TagConfusion2}. It can be observed that as the length of the frozen tag increases, both the average and maximum accumulated noise powers increase. This suggests that increasing the length of the frozen tag makes it more difficult for Eve to estimate the raw tags. %Notably, the minimum accumulated noise power shown in the figure is always four times greater than the receiver noise power, as derived from the coefficient in (\ref{Sigmaw}).

For the spoofing attack, we simulate the normalized symmetric difference between the positions of the frozen tag for Eve and Alice as shown in Figure \ref{PerSD}, where positions of Alice's frozen tag are generated by the secret key, while those of Eve are random due to his lack of knowledge of the key. The following observations can be made. First, as the length of the message $N$ increases, the normalized symmetric difference gradually increases. This is because the overlap probability between Eve's and Alice's frozen tag decreases with an increasing $N$. Second, as the length of the frozen tag  $N_\text{e}$ increases, the normalized symmetric difference decreases. %This suggests that increasing the length of the message while reducing the length of the frozen tag can degrade the overlap between positions of Eve's and Alice's frozen tags.

\vspace*{-0.3cm}
\subsection{Compatibility Evaluation}
\vspace*{-0.3cm}

\begin{figure*}[t!]
	\centering
	\begin{minipage}[t]{0.49\linewidth}
		\centering
		\includegraphics[width=3in]{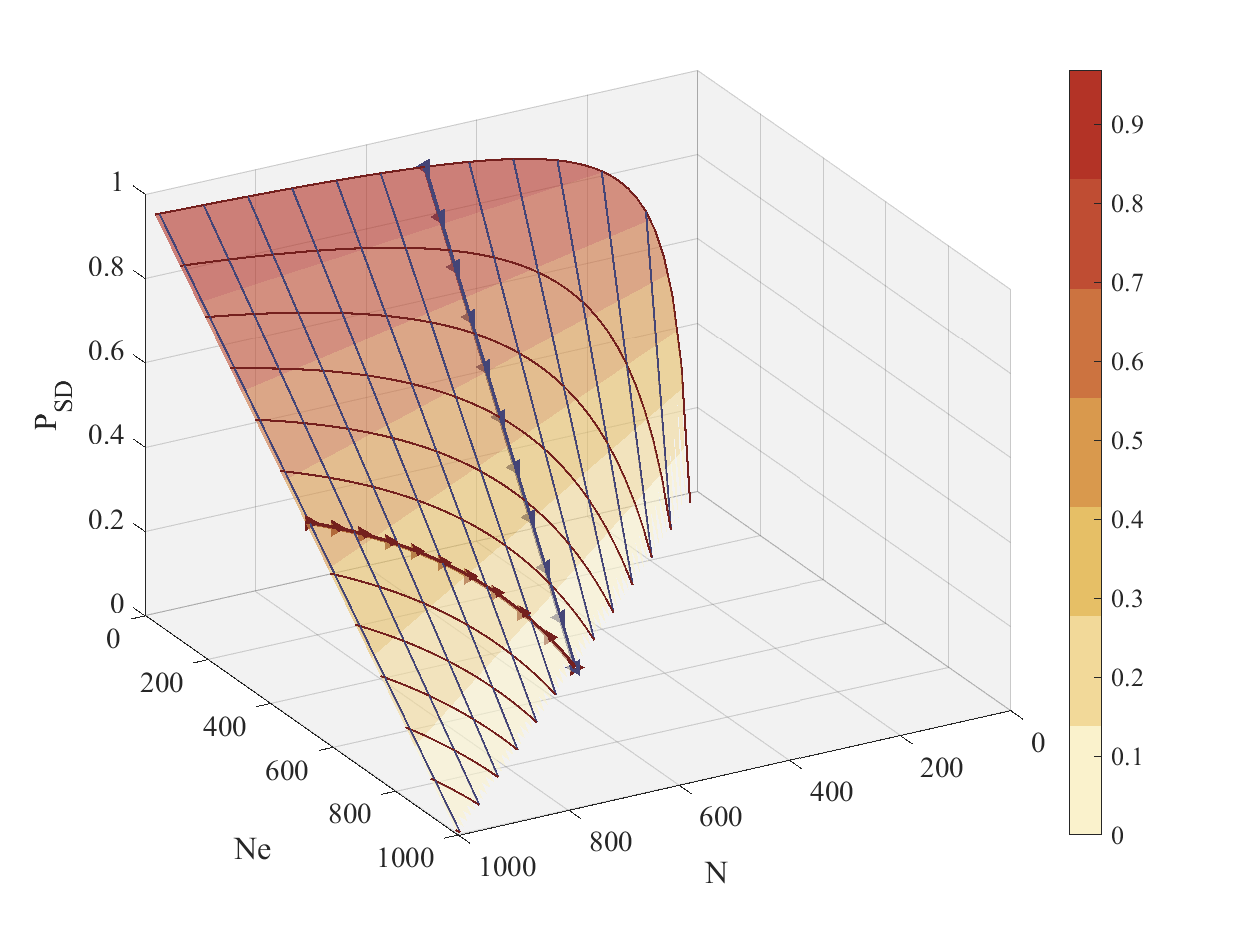}
		\caption{Variation of the normalized symmetric difference between the positions of the frozen tags for Eve and Alice versus lengths of  message $N$ and frozen tag $N_\text{e}$, where an increasing normalized symmetric difference indicates a lower degree of overlap in the positions of the frozen tag.}
		\label{PerSD}
	\end{minipage}
	\hfill
	\begin{minipage}[t]{0.49\linewidth}
		\centering
		\includegraphics[width=2.35in]{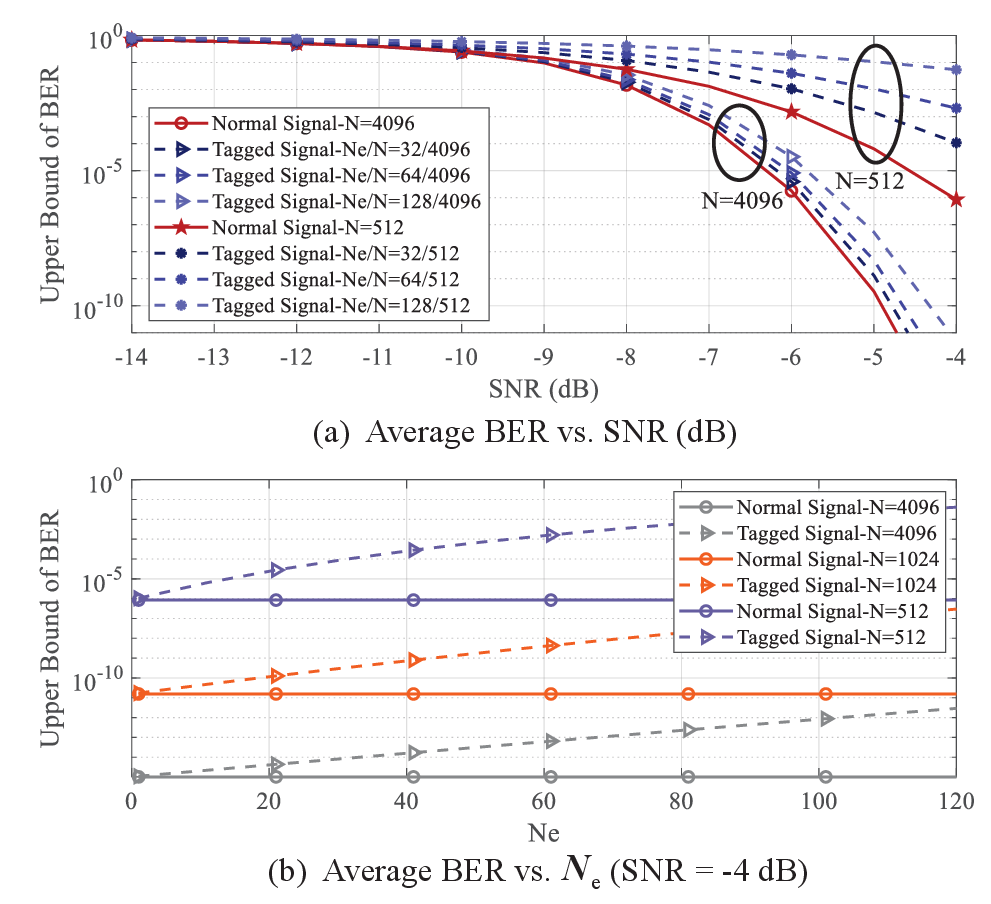}
		\caption{A comparison of the upper bounds of BER for tagged messages and normal messages ($N_\text{e}=0$), including (a) the variation of BER with SNR and (b) the variation of BER with different lengths of the frozen tags.}
		\label{BERab}
	\end{minipage}
	\vspace*{-0.3cm}
\end{figure*}

\begin{observation}
	\textit{The BER of the tagged signal tends toward that of the normal signal with increasing length of the message and decreasing length of the frozen tag. (cf. Figure \ref{BERab})}
\end{observation}
\vspace*{-0.2cm}

\noindent We simulate the upper bounds of the BER for messages with and without inserted frozen tags to demonstrate the compatibility of the proposed framework as shown in Figure \ref{BERab}. It is noteworthy that, due to the unawareness of the unconscious receiver for the authentication, the legitimate parties should select the reliable sub-channels $\mathcal{I}$ without considering the frozen tags for encoding and decoding the message to facilitate decoding by the unaware receiver. From Figure \ref{BERab}(a), we can observe the following phenomena. First, the BER decreases as the SNR increases. Second, the BER increases with longer frozen tags. However, the increase in BER is less pronounced  with a longer message. This is because the insertion of frozen tags interferes with the decoding of the message. When the message length is large, the proportion of the frozen tag in the message is lower, thus its impact on the message is less significant. From Figure \ref{BERab}(b), we can observe that as the frozen tag length increases, the BER deteriorates progressively. However, when the message length is longer, the upper bound of BER for the message remains at a lower level, indicating that the proposed scheme exhibits high compatibility. 
%\subsection{Discussion}
%The simulation results reveal that the robustness, compatibility, and security of the proposed framework are jointly governed by the lengths of the frozen tag, the anchor information, and the message. For instance, a longer frozen-tag enhances robustness at the expense of reduced security and compatibility. Thus, these parameters should be tailored based on the specific requirements in practical deployments. Furthermore, in scenarios demanding high compatibility, the impact of frozen tags can be distributed by embedding them across multiple messages. Conversely, for applications prioritizing robustness, a dual-factor authentication mechanism can be formed by integrating the proposed frozen tag-based framework with an uncoded tag-based approach. The detailed design of these extensions is beyond the scope of this paper.

\vspace*{-0.5cm}
\section{Conclusion}\label{6}
\vspace*{-0.2cm}
To address the issues of authentication performance degradation caused by user interference and security vulnerabilities due to direct tag embedding in traditional PLA frameworks, this paper proposes a novel PLA framework based on frozen tags. In this framework, a frozen tag, generated by both anchor information and the raw tag, is randomly inserted into the message. Moreover, an analysis of the proposed framework is conducted in terms of the robustness, security and compatibility. Specifically, for security, we analyze the framework's resilience against eavesdropping and spoofing attacks by Eve, indicating that it is difficult for Eve to estimate both the raw tag and the positions of the frozen tags. For robustness and compatibility, we derive a union bound on the authentication probability and analyze the BER of the message, respectively. Theoretical analysis and simulation demonstrate that the proposed framework significantly enhances both detection performance and security.

%\clearpage

%\Acknowledgements{This work was supported by National Natural Science Foundation of China under Grant 62171445 and Grant 62201590.}

%\footnotetext[1]{Footnote 1.}
%\footnotetext[2]{Footnote 2.}

%%%%%%%%%%%%%%%%%%%%%%%%%%%%%%%%%%%%%%%%%%%%%%%%%%%%%%%
%%% Acknowledgements. 致谢
%%%%%%%%%%%%%%%%%%%%%%%%%%%%%%%%%%%%%%%%%%%%%%%%%%%%%%%

%%%%%%%%%%%%%%%%%%%%%%%%%%%%%%%%%%%%%%%%%%%%%%%%%%%%%%%
%%% Supplements. 补充材料, 非必选
%%%%%%%%%%%%%%%%%%%%%%%%%%%%%%%%%%%%%%%%%%%%%%%%%%%%%%%

%%%%%%%%%%%%%%%%%%%%%%%%%%%%%%%%%%%%%%%%%%%%%%%%%%%%%%%
%%% Open Access Funding Note. 开放获取基金来源说明
%%%%%%%%%%%%%%%%%%%%%%%%%%%%%%%%%%%%%%%%%%%%%%%%%%%%%%%
%\FundingNote{\textbf{Open Access} funding enabled and organized by CAUL and its Member Institutions.}

%%%%%%%%%%%%%%%%%%%%%%%%%%%%%%%%%%%%%%%%%%%%%%%%%%%%%%%
%%% Reference section. 参考文献
%%% Citation in the content using "some words~\cite{1,2}".
%%% ~ is needed to make the reference number is on the same line with the word before it.
%%% Using scis.bst to format the style if the ref.bib file is included, e.g.,
%%% \bibliographystyle{scis}
%%% \bibliography{ref}
%%%%%%%%%%%%%%%%%%%%%%%%%%%%%%%%%%%%%%%%%%%%%%%%%%%%%%%
\bibliographystyle{scis}
\bibliography{ylbib}

\begin{thebibliography}{10}
\providecommand{\url}[1]{\texttt{#1}}
\providecommand{\urlprefix}{URL }
\providecommand{\bibinfo}[2]{#2}

\bibitem{iot}
Aouedi O, Vu T~H, Sacco A,  et~al.
\newblock A survey on intelligent {Internet of Things}: applications, security,
  privacy, and future directions.
\newblock IEEE Commun. Surv. Tutor., 2025, 27: 1238--1292

\bibitem{6G}
Jiang W, Han B, Habibi M~A,  et~al.
\newblock The road towards 6{G}: a comprehensive survey.
\newblock IEEE Open J. Commun. Soc., 2021, 2: 334--366

\bibitem{indusIOT}
Chi H~R, Wu C~K, Huang N~F,  et~al.
\newblock A survey of network automation for industrial {Internet-of-Things}
  toward industry 5.0.
\newblock IEEE Trans. Ind. Informat., 2023, 19: 2065--2077

\bibitem{IOV}
Li C, Dong M, Fu Y,  et~al.
\newblock Integrated sensing, communication, and computation for {IoV}:
  challenges and opportunities.
\newblock IEEE Commun. Surv. Tutor., 2025, 28: 1136--1168

\bibitem{IOTT}
Illi E, Qaraqe M, Althunibat S,  et~al.
\newblock Physical layer security for authentication, confidentiality, and
  malicious node detection: a paradigm shift in securing {IoT} networks.
\newblock IEEE Commun. Surv. Tutor., 2024, 26: 347--388

\bibitem{ULA}
Zhou Y, Cao L, Qiao Z,  et~al.
\newblock An efficient identity authentication scheme with dynamic anonymity
  for vanets.
\newblock IEEE Internet Things J., 2023, 10: 10052--10065

\bibitem{XieSurvey}
Xie N, Li Z,  Tan H.
\newblock A survey of physical-layer authentication in wireless communications.
\newblock IEEE Commun. Surv. Tutor., 2021, 23: 282--310

\bibitem{XieTSP}
Zhang Z, Tan H, Xie N,  et~al.
\newblock Jamming detection based on source enumeration in massive channel
  systems.
\newblock IEEE Trans. Signal Process., 2026, 74: 1263--1276

\bibitem{Channel1}
Sheng Y, Tan K, Chen G,  et~al.
\newblock Detecting 802.11 {MAC} layer spoofing using received signal strength.
\newblock In: Proceedings of Conf. Comput. Commun. (INFOCOM), 2008.
\newblock 1768-1776

\bibitem{Channel2}
Liu F~J, Wang X,  Primak S~L.
\newblock A two dimensional quantization algorithm for {CIR}-based physical
  layer authentication.
\newblock In: Proceedings of Int. Conf. Commun. (ICC), 2013.
\newblock 4724-4728

\bibitem{Channel3}
Xiao L, Greenstein L~J, Mandayam N~B,  et~al.
\newblock Channel-based spoofing detection in frequency-selective rayleigh
  channels.
\newblock IEEE Trans. Wireless Commun., 2009, 8: 5948--5956

\bibitem{device1}
Hao P, Wang X,  Behnad A.
\newblock Performance enhancement of {I/Q} imbalance based wireless device
  authentication through collaboration of multiple receivers.
\newblock In: Proceedings of Int. Conf. Commun. (ICC), 2014.
\newblock 939-944

\bibitem{PLA2008}
Yu P~L, Baras J~S,  Sadler B~M.
\newblock Physical-layer authentication.
\newblock IEEE Trans. Inf. Forensic Secur., 2008, 3: 38--51

\bibitem{XieSlope}
Xie N,  Chen C.
\newblock Slope authentication at the physical layer.
\newblock IEEE Trans. Inf. Forensic Secur., 2018, 13: 1579--1594

\bibitem{ITS}
Maurer U.
\newblock Authentication theory and hypothesis testing.
\newblock IEEE Trans. Inf. Theory, 2000, 46: 1350--1356

\bibitem{emc}
Kokuvi Angélo~Passah A, Chorti A,  de~Lamare R~C.
\newblock Enhanced multiuser {CSI}-based physical layer authentication based on
  information reconciliation.
\newblock IEEE Wireless Commun. Lett., 2025, 14: 544--548

\bibitem{CIR1}
Tugnait J~K,  Kim H.
\newblock A channel-based hypothesis testing approach to enhance user
  authentication in wireless networks.
\newblock In: Proceedings of Int. Conf. Commun. Syst. Netw. (COMSNETS), 2010.
\newblock 1-9

\bibitem{CFO1}
Hou W, Wang X,  Chouinard J~Y.
\newblock Physical layer authentication in {OFDM} systems based on hypothesis
  testing of {CFO} estimates.
\newblock In: Proceedings of Int. Conf. Commun. (ICC), 2012.
\newblock 3559-3563

\bibitem{CFO2}
Hou W, Wang X, Chouinard J~Y,  et~al.
\newblock Physical layer authentication for mobile systems with time-varying
  carrier frequency offsets.
\newblock IEEE Trans. Commun., 2014, 62: 1658--1667

\bibitem{BTP}
Wang C, Sha M, Xiong W,  et~al.
\newblock Blind tag-based physical-layer authentication.
\newblock IEEE-ACM Trans. Netw., 2024, 32: 1--14

\bibitem{CR}
Koorapaty H, Hassan A,  Chennakeshu S.
\newblock Secure information transmission for mobile radio.
\newblock IEEE Commun. Lett., 2000, 4: 52--55

\bibitem{CRH}
Xie N, Zhang J, Zhang Q,  et~al.
\newblock Hybrid physical-layer authentication.
\newblock IEEE. Trans. Mob. Comput., 2024, 23: 1295--1311

\bibitem{WFRFT}
Zhang N, Fang X, Wang Y,  et~al.
\newblock Physical-layer authentication for {Internet of Things} via
  {WFRFT}-based {Gaussian} tag embedding.
\newblock IEEE Internet Things J., 2020, 7: 9001--9010

\bibitem{ATBS}
Tan H, Xu Y, Du J,  et~al.
\newblock Asynchronous tag-based physical-layer authentication in wireless
  communications.
\newblock IEEE Trans. Wireless Commun., 2025, 24: 7809--7821

\bibitem{PhaPLA}
Xie N, Xiong W, Chen J,  et~al.
\newblock Multiple phase noises physical-layer authentication.
\newblock IEEE Trans. Commun., 2022, 70: 6196--6211

\bibitem{Interf2}
Bariah L, Muhaidat S,  Al-Dweik A.
\newblock Error probability analysis of non-orthogonal multiple access over
  nakagami-$m$ fading channels.
\newblock IEEE Trans. Commun., 2019, 67: 1586--1599

\bibitem{Interf1}
Moshavi S.
\newblock Multi-user detection for {DS}-{CDMA} communications.
\newblock IEEE Commun. Mag., 1996, 34: 124--136

\bibitem{MulOb}
Cai Y, Wang W, Chen Y,  et~al.
\newblock Multiple cooperative attackers for tag-based physical layer
  authentication.
\newblock IEEE Commun. Mag., 2023, 61: 165--171

\bibitem{XiePSA}
Xie N, Chen C,  Ming Z.
\newblock Security model of authentication at the physical layer and
  performance analysis over fading channels.
\newblock IEEE Trans. Dependable Secur. Comput., 2021, 18: 253--268

\bibitem{LDPC}
Gallager R.
\newblock Low-density parity-check codes.
\newblock IEEE Trans. Inf. Theory, 1962, 8: 21--28

\bibitem{Turbo}
Berrou C, Glavieux A,  Thitimajshima P.
\newblock Near shannon limit error-correcting coding and decoding: Turbo-codes.
\newblock In: Proceedings of IEEE Int. Conf. Communications, 1993.
\newblock 1064-1070

\bibitem{Polar}
Arikan E.
\newblock Channel polarization: A method for constructing capacity-achieving
  codes.
\newblock In: Proceedings of IEEE Int. Symp. Inf. Theory (ISIT), 2008.
\newblock 1173-1177

\bibitem{PDUB}
Tal I,  Vardy A.
\newblock How to construct polar codes.
\newblock IEEE Trans. Inf. Theory, 2013, 59: 6562--6582

\bibitem{GA}
Wu D, Li Y,  Sun Y.
\newblock Construction and block error rate analysis of polar codes over {AWGN}
  channel based on gaussian approximation.
\newblock IEEE Commun. Lett., 2014, 18: 1099--1102

\bibitem{SD}
Lopez-Martinez F~J,  Romero-Jerez J~M.
\newblock Asymptotically exact approximations for the symmetric difference of
  generalized {Marcum} $q$-functions.
\newblock IEEE Trans. Veh. Techn., 2015, 64: 2154--2159

\end{thebibliography}
%\begin{thebibliography}{99}

%\bibitem{ASGMIN} Du M, Yang P, Liu Y Q, et al. HYShield: a multi-layered security framework for air-space-ground-maritime networks in 6G scenarios. IEEE Commun Mag, 2025, 63: 18--24

%\bibitem{TWC22Fang} Fang X R, Feng W, Wang Y M, et al. NOMA-based hybrid satellite-UAV-terrestrial networks for 6G maritime coverage. IEEE Trans Wireless Commun, 2023, 22: 138--152

%\bibitem{TITS22Yu} Yu P, Ding Y H, Li Z F, et al. Energy-efficient coverage and capacity enhancement with intelligent UAV-BSs deployment in 6G edge networks. IEEE Trans Intell Transp Syst, 2023, 24: 7664--7675

%\bibitem{1} Author A, Author B, Author C. Reference title. Journal, 2024, 38: 13--28

%\bibitem{2} Author A, Author B, Author C, et al. Reference title. In: Proceedings of Conference, Place, 2024. 6--12

%\end{thebibliography}

%%%%%%%%%%%%%%%%%%%%%%%%%%%%%%%%%%%%%%%%%%%%%%%%%%%%%%%
%%% Appendix sections. 附录章节, 非必选
%%%%%%%%%%%%%%%%%%%%%%%%%%%%%%%%%%%%%%%%%%%%%%%%%%%%%%%
%\begin{appendix}
%\section{Name}

%\end{appendix}

\end{document}